\documentclass[%
reprint,
superscriptaddress,
nofootinbib,
amsmath,amssymb,
aps,
]{revtex4-2}

\usepackage{graphicx}
\usepackage{dcolumn}
\usepackage{bm}
\usepackage{float}
\usepackage[caption=false]{subfig}
\usepackage[hidelinks]{hyperref}

\begin{document}

\preprint{APS/123-QED}

\title{Exclusive $J/\psi$ photoproduction in ultraperipheral Pb+Pb collisions at the LHC to next-to-leading order perturbative QCD}

\author{K.~J.~Eskola}

\affiliation{%
University of Jyvaskyla, Department of Physics, P.O. Box 35, FI-40014 University of Jyvaskyla, Finland}%
\affiliation{Helsinki Institute of Physics, P.O. Box 64, FI-00014 University of Helsinki, Finland}

\author{C.~A.~Flett}%

\affiliation{%
University of Jyvaskyla, Department of Physics, P.O. Box 35, FI-40014 University of Jyvaskyla, Finland}%
\affiliation{Helsinki Institute of Physics, P.O. Box 64, FI-00014 University of Helsinki, Finland}
 

 \author{V.~Guzey, T.~L\"oyt\"ainen}
 \email{topi.m.o.loytainen@jyu.fi}
 
 \affiliation{%
University of Jyvaskyla, Department of Physics, P.O. Box 35, FI-40014 University of Jyvaskyla, Finland}%
\affiliation{Helsinki Institute of Physics, P.O. Box 64, FI-00014 University of Helsinki, Finland}
 
 \author{H.~Paukkunen}
 
\affiliation{%
University of Jyvaskyla, Department of Physics, P.O. Box 35, FI-40014 University of Jyvaskyla, Finland}%
\affiliation{Helsinki Institute of Physics, P.O. Box 64, FI-00014 University of Helsinki, Finland}


\date{\today}

\begin{abstract}
We present the first next-to-leading-order (NLO) perturbative QCD (pQCD) study of rapidity-differential cross sections of coherent exclusive photoproduction of $J/\psi$ mesons in heavy-ion ultraperipheral collisions (UPCs) at the LHC, $d\sigma/dy(\text{Pb} + \text{Pb} \rightarrow \text{Pb} + J/\psi + \text{Pb})$. For this, we account for the photon-nucleon NLO cross sections at the forward limit, the $t$ dependence using a standard nuclear form factor, and the photon fluxes of the colliding nuclei. Approximating the generalized parton distributions with their forward-limit parton distribution functions (PDFs), we quantify the NLO contributions in the cross sections, show that the real part of the amplitude and quark-PDF contributions must not be neglected, quantify the uncertainties arising from the scale-choice and PDFs, and compare our results with ALICE, CMS and LHCb $J/\psi$ photoproduction data in Pb+Pb UPCs, exclusive $J/\psi$ photoproduction data from HERA, and LHCb data in p+p. The scale dependence in $d\sigma/dy(\text{Pb} + \text{Pb} \rightarrow \text{Pb} + J/\psi + \text{Pb})$ is significant, but we can find a scale-choice that reproduces the Pb+Pb UPC data both at 2.76 and 5.02 TeV collision energies. This process has traditionally been suggested to be a direct probe of nuclear gluon distributions. We show that the situation changes rather dramatically from LO to NLO: the NLO cross sections reflect the nuclear effects of both gluons and quarks in a complicated manner where the relative signs of the LO and NLO terms in the amplitude play a significant role. 
\end{abstract}

\maketitle

\section{\label{Sec:Intro}Introduction}
Ultraperipheral collisions (UPCs) are collisions of hadrons or nuclei which take place at large impact parameters in such a way that only the electromagnetic field of one of the colliding particles interacts with the other particle \cite{Bertulani:1987tz,Nystrand:2006gi,Baltz:2007kq}. Coherent photoproduction of $J/\psi$ heavy vector-mesons in UPCs of Lead nuclei at the CERN Large Hadron Collider (LHC), the exclusive process 
$\text{Pb} + \text{Pb} \rightarrow \text{Pb} + J/\psi + \text{Pb}$,
has been suggested to be an efficient direct probe of collinear nuclear gluon distributions, $g_\text{Pb}(x,Q^2)$, at factorization scales of the order of the vector-meson mass, $Q^2 = \mathcal{O}( M_V^2)$, and small longitudinal-momentum fractions $x = \mathcal{O} (M_{V}^2/W^2)$, where $W$ is the photon-nucleon center-of-momentum-system (c.m.s.) energy 
\cite{Adeluyi:2011rt,Adeluyi:2012ph,Guzey:2016piu,Guzey:2020ntc,Guzey:2013xba,Guzey:2013qza,Guzey:2016qwo,Jones:2016ldq}. This exciting possibility derives from the fact that in such an exclusive  process of no hadronic activity,  one of the colliding nuclei serves as a source of equivalent real Weizsäcker-Williams photons which probe a color-singlet gluon- or quark-initiated ladder from the other nucleus via formation of a heavy quark-antiquark pair. As first discussed by Ryskin in Ref.~\cite{Ryskin:1992ui} in the context of the free-proton process $\gamma+\text{p}\rightarrow J/\psi + \text{p}$, in the leading order (LO) perturbative QCD (pQCD) only the gluon-ladder processes contribute, and neglecting the longitudinal-momentum imbalance (skewedness) in the ladder and the subleading real part of the amplitude, the forward scattering amplitude factorizes into a calculable hard part and $g_\text{p}(x,Q^2)$. Thus the cross section of $J/\psi$ becomes proportional to  $[g_\text{p}(x,Q^2)]^2$, making the process a very promising one for probing the gluon distribution. This idea has then been transferred to ultraperipheral nucleus-nucleus collisions (UPCs) in e.g. Refs.~\cite{Adeluyi:2011rt,Adeluyi:2012ph}. Also Monte Carlo event simulations of this process in the UPCs have been developed, such as STARlight~\cite{Klein:2016yzr} and SuperChic~\cite{Harland-Lang:2020veo}. Exclusive photoproduction of $J/\psi$ has also been widely studied in the dipole picture, especially in the high-energy Color-Glass-Condensate approximation of QCD, see e.g. Refs.~\cite{Goncalves:2005yr,Lappi:2013am,Mantysaari:2017dwh,Mantysaari:2017slo,Cepila:2017nef,Mantysaari:2019jhh,Sambasivam:2019gdd,Klein:2019qfb,Caldwell:2010zza,Bendova:2020hbb,Mantysaari:2021ryb}. 

With the experimental data being released from the LHC, the situation is becoming ever more interesting. Firstly, the exclusive coherent $J/\psi$ photoproduction cross sections involving real photons have been measured in electron-proton collisions at the DESY-HERA collider by the H1 \cite{H1:2000kis} and ZEUS \cite{ZEUS:2002wfj} collaborations, and extracted also from the LHCb measurements of the process $\text{p}+\text{p}\rightarrow \text{p} + J/\psi + \text{p}$ at the LHC \cite{LHCb:2014acg,LHCb:2018rcm}. For detailed NLO pQCD studies of these, see e.g. Refs. \cite{Ivanov:2004vd,Jones:2015nna,Flett:2019nga,Flett:2019pux,Flett:2020duk,Flett:2021xsl}. From the viewpoint of the UPCs, these data sets offer also an importantly long lever arm in the photon-proton c.m.s. energy $W$ for cross-checking the pQCD calculations and understanding the necessary modeling input. Secondly, in Pb+Pb UPCs at the LHC, the ALICE collaboration has measured the rapidity-differential cross section of $\text{Pb} + \text{Pb} \rightarrow \text{Pb} + J/\psi + \text{Pb}$ both at midrapidity \cite{ALICE:2021gpt,ALICE:2013wjo} and at forward/backward rapidities \cite{ALICE:2019tqa,ALICE:2012yye} at nucleon-nucleon c.m.s. energies $\sqrt{s_{\rm NN}}=5.02$ TeV  and 2.76 TeV. The CMS collaboration has performed the corresponding measurement at $\sqrt{s_{\rm NN}}=2.76$ TeV in one off-central rapidity bin that lies conveniently just between the ALICE rapidity bins \cite{CMS:2016itn}. The LHCb collaboration has recently released their 5.02 TeV data at forward/backward rapidities \cite{LHCb:2021bfl}, overlapping with the ALICE rapidity region.  Very interestingly, however, the ALICE and LHCb forward/backward-rapidity 5.02 TeV data sets do not seem to be fully compatible with each other, which clearly calls for further analyses.

Until now, exclusive $J/\psi$ photoproduction in ultraperipheral nuclear collisions has been studied only to LO pQCD. Now that the LHC experiments are measuring these cross sections to an increasing accuracy, and hopefully also for other UPC systems than Pb+Pb in the future \cite{Brewer:2021kiv}, it is clearly of high priority to extend the theory calculations to NLO pQCD. In particular we wish to study whether/how this process could be included in the global analyses of nuclear PDFs, such as in Refs.~\cite{Kovarik:2015cma, Eskola:2016oht,AbdulKhalek:2020yuc,Eskola:2021nhw,Khalek:2022zqe}, in the future. These are the main motivations for our present NLO study. Also interestingly, so far the LO pQCD, or dipole picture, calculations have not been able to reproduce simultaneously the mid- and forward/backward-rapidity data, see e.g.~\cite{ALICE:2021gpt}. This, together with the mentioned incompatibility between the LHCb and ALICE data, serves also as further motivation for our current NLO pQCD study. 

The NLO pQCD calculation of cross sections for exclusive photoproduction of heavy vector mesons $V$ off the free proton,  $\sigma(\gamma+\text{p}\rightarrow V + \text{p})$, using collinear factorization at the amplitude level, has been performed first by Ivanov et al. in Ref.~\cite{Ivanov:2004vd}, followed then by other groups in Refs.~\cite{Jones:2013pga,Jones:2015nna,Chen:2019uit,Flett:2021xsl,Flett:2021ghh}. To be exact, collinear factorization here refers to the factorization of the amplitude to calculable NLO pQCD pieces and to the generalized parton distributions (GPDs) \cite{Collins:1996fb} which at the forward limit relax into the usual PDFs~\cite{Diehl:2003ny}. If such a limit is not assumed, then the GPDs have to be modeled in some way, e.g. as suggested in Refs.~\cite{Shuvaev:1999ce,Shuvaev:1999fm,Freund:2002qf,Martin:2008gqx,Kumericki:2009uq,Harland-Lang:2013xba,Constantinou:2020hdm}. As is shown already in Ref.~\cite{Ivanov:2004vd}, the full NLO calculation of coherent exclusive photoproduction of $J/\psi$ mesons in $\gamma+\text{p}$ collisions, which includes both the imaginary and real parts of the amplitude precisely as they are, and assumes a certain model for the gluon- and quark-GPDs \cite{Freund:2002qf},  depends rather heavily on the choice of the renormalization/factorization scale,  $Q={\cal O}(M_{J/\psi})$, while for the photoproduction of $\Upsilon$ mesons, which probes a higher scale $Q={\cal O}(M_{\Upsilon})$, the situation improves somewhat. Discussion of a systematic procedure for diminishing the scale dependence in the NLO calculation of exclusive $J/\psi$ photoproduction in $\gamma+\text{p}$  collisions can be found in \cite{Flett:2019nga,Flett:2019pux,Flett:2020duk,Flett:2021xsl}, but in the present exploratory NLO study for the nuclear UPCs we do not follow this avenue.

In the current paper, we present the first NLO pQCD study of exclusive photoproduction of $J/\psi$ mesons in ultraperipheral Pb+Pb collisions at the LHC, with collinear factorization at the amplitude level. Exploiting the analytic results of the impressive calculation of Ref.~\cite{Ivanov:2004vd}, we have built a numerical code of our own for the rapidity-differential $J/\psi$ photoproduction UPC cross sections, $d\sigma/dy(\text{Pb} + \text{Pb} \rightarrow \text{Pb} + J/\psi + \text{Pb})$. These consist of a rather non-trivial numerical evaluation of the  differential NLO forward  photoproduction  cross sections $d\sigma/dt(\gamma +\text{Pb} \rightarrow J/\psi + \text{Pb})$ at vanishing Mandelstam variable $t$ based on Ref.~\cite{Ivanov:2004vd}, supplemented with a straightforward computation of the nuclear form factor to account for the $t$ dependence of the cross section, as well as a non-trivial numerical evaluation of the photon fluxes from the colliding Lead nuclei based on Refs.~\cite{Vidovic:1992ik,Guzey:2013taa}. In the current exploratory NLO study we adopt the simplest possible, forward-limit, approximation for the GPDs where they become just the usual PDFs. With such a ``bare bones" GPD/PDF NLO framework, our goal is to test as transparently as possible, and without any additional normalization factors (which typically appear in LO studies) or modeling, how directly and efficiently the exclusive photoproduction of $J/\psi$ mesons in Pb+Pb UPCs at the LHC actually probes the nuclear gluon distributions. 

In what follows, we will first chart the scale dependence of the NLO cross sections, and compare the situation with the LO case, too. Even though the scale dependence of the NLO cross sections is known to be quite strong \cite{Ivanov:2004vd}, we will show that interestingly a reasonable ``optimal" scale choice can be found, with which we can, perhaps contrary to our initial expectations, simultaneously reproduce the 5.02 TeV ALICE mid-rapidity \cite{ALICE:2021gpt} and the LHCb forward-rapidity \cite{LHCb:2021bfl} data, and also the 2.76 TeV ALICE \cite{ALICE:2013wjo,ALICE:2012yye} and CMS \cite{CMS:2016itn} data. We will also study the corresponding NLO cross sections in photon-proton collisions, as well as their scale dependence, against the HERA and LHCb data.

We will also break down the NLO calculation into the contributions from the imaginary and real parts, as well from the gluon and quark PDFs, and show (in accordance with Ref. \cite{Ivanov:2004vd}) that the real part of the amplitude as well as the quark contributions both have a sizeable contribution and hence must not be neglected. This result indicates that, contrary to what is often claimed based on the LO results, exclusive $J/\psi$ photoproduction in UPCs is not as direct a probe of the gluon distributions as perhaps previously thought. We will chart, by comparing the predictions obtained with the EPPS16 \cite{Eskola:2016oht} nuclear PDFs and CT14NLO free proton PDFs \cite{Dulat:2015mca}, and nCTEQ15 \cite{Kovarik:2015cma} and nNNPDF2.0 \cite{AbdulKhalek:2020yuc} nuclear PDFs, how the gluon and quark PDFs manifest themselves in the $J/\psi$ photoproduction UPC cross sections at different rapidities. In particular, using EPPS16, we will show that the manifestation of the nuclear effects is non-trivial and influenced especially by the relative signs of the different contributions in the amplitude. Finally, as one of the main goals of the paper, we will study how the uncertainties of the nuclear and free-proton PDFs propagate into the $J/\psi$ photoproduction UPC cross sections. 

The rest of this paper will proceed as follows: To make our study more accessible especially for the heavy-ion community and non-GPD-experts in general, we will recapitulate the theoretical NLO framework with collinear factorization and GPDs/PDFs in Sec.~2. Also the calculation of the photon fluxes and evaluation of the necessary nuclear form factors are presented there. The main results of the paper, the numerical evaluation of the  coherent exclusive $J/\psi$ photoproduction cross sections in Pb+Pb UPCs at the LHC, their analysis and comparison with the experimental data, are presented in Sec.~3. Finally, a discussion and outlook are given in Sec.~4. 

\section{\label{Sec:TheoFrame}Theoretical Framework}

\subsection{Differential Cross Section}

In this section, we recapitulate the theoretical framework we use in our calculations for the exclusive process 
\begin{equation}
A_1(p_1) + A_2(p_2) \rightarrow A_1(p'_1) + V(p'_3) + A_2(p'_2) \,, \nonumber
\end{equation}
where $A_{1,2}$ denote the colliding nuclei and $V$ is some vector meson (in this paper $V=J/\psi$). The initial-state momenta are labeled by $p_i$ and the final-state momenta by $p'_i$. Within the equivalent-photon (Weizs\"acker-Williams) approximation \cite{Drees:1988pp,Bertulani:2005ru,Baltz:2007kq}, the total cross section can be expressed as
\begin{equation}
\begin{split}
    \sigma^{A_1 A_2\rightarrow A_1VA_2} = &\int dk^+ \frac{dN_{\gamma}^{A_1}(k^+)}{dk^+} \sigma^{\gamma(k^+) A_2 \rightarrow VA_2} \\
    + & \int dk^- \frac{dN_{\gamma}^{A_2}(k^-)}{dk^-} \sigma^{A_1\gamma(k^-) \rightarrow A_1 V} , \label{eq:xsec2}
\end{split}
\end{equation}
where $dN_{\gamma}^{A_i}(k)/dk$ is the centrality-integrated distribution (or flux) of photons from the nucleus $A_i$ as a function of photon energy $k$, and $\sigma^{\gamma(k^+) A_2 \rightarrow VA_2}$ and $\sigma^{A_1\gamma(k^-) \rightarrow VA_1}$ are the cross sections for the photoproduction processes
\begin{align}
\gamma(k_1) + A_2(p_2) & \rightarrow V(p'_3) + A_2(p'_2) \nonumber \,,  \\[4pt] 
A_1(p_1) + \gamma(k_2) & \rightarrow A_1(p'_1) + V(p'_3)  \nonumber \,.         
\end{align}
In the equivalent-photon approximation the photon momenta $k_{1,2}$ are considered to be collinear with colliding nuclei, and $|\vec k_{1,2}|=k^\pm$. The experimental data in Pb-Pb collisions \cite{CMS:2016itn,ALICE:2019tqa,ALICE:2021gpt,LHCb:2021bfl} are differential with respect to the rapidity $y$ of the vector meson. At fixed rapidity and transverse momentum $p_{\rm T}$ of produced vector meson, the photon momentum can be expressed as
\begin{align}
k^\pm = \frac{M_V^2 - t}{2M_{\rm T}e^{\mp y}} \,, 
\end{align}
where $t$ refers to the square of the momentum transferred to the target nucleus, $t=(k_{1,2}-p'_3)^2$, and $M_{\rm T} = \sqrt{M_V^2 + p_{\rm T}^2}$ is the transverse mass. In the typical case $|t| \ll M_V^2$ and $p^2_{\rm T} \ll M_V^2$ (see e.g. Ref.~\cite{ALICE:2021tyx}) so that to a very good approximation
\begin{align}
k^\pm \approx \frac{M_Ve^{\pm y}}{2} \,. 
\end{align}
It then follows that 
\begin{equation}
   \begin{split}
        \frac{d\sigma^{A_1A_2\rightarrow A_1VA_2} }{dy} = &\left[ k \frac{dN_\gamma^{A_1} (k)}{dk} \sigma^{\gamma(k) A_2 \rightarrow VA_2} \right]_{k=k^+} \\
        + &\left[ k \frac{dN_\gamma^{A_2} (k)}{dk} \sigma^{A_1 \gamma(k) \rightarrow A_1 V}  \right]_{k=k^-} \,. 
   \end{split}\label{XS_plus_minus}
\end{equation}

\subsection{Photoproduction Cross Section}
We will assume that the invariant matrix element $\mathcal{M}^{\gamma A \rightarrow VA}$ for the photoproduction process can be factored into two parts, the matrix element evaluated at $t=0$ and a nuclear form factor $F_A(t)$ (also called the two-gluon form factor~\cite{Ryskin:1992ui}) \cite{Klein:1999qj},
\begin{equation}
    \mathcal{M}^{\gamma A \rightarrow VA} (W,t) = \mathcal{M}^{\gamma N \rightarrow VN}_A(W,0) F_A (t) \,,
\end{equation}
where $N$ labels a bound nucleon and $W$ is the c.m.s. energy of the photon-nucleon collision. It follows that the photoproduction cross section then becomes
\begin{align}
\sigma^{\gamma A \rightarrow VA} (W) & = \frac{d\sigma_A^{\gamma N \rightarrow VN}}{dt} \bigg|_{t=0} \int\limits_{t_{\rm min}}^\infty dt' |F_A(-t')|^2, \label{eq:xsec1} \\
\frac{d\sigma_A^{\gamma N \rightarrow VN}}{dt} & = \frac{\overline{|\mathcal{M}^{\gamma N \rightarrow VN}_A|^2}}{16\pi W^4} \,,
\end{align}
where $\overline{|\mathcal{M}^{\gamma N \rightarrow VN}_A|^2}$ is the square of the per-nucleon matrix element averaged (summed) over the initial-state (final-state) polarizations. The minimum momentum transfer squared is given by $t_{\rm min} = [M_V^2/(4k\gamma_{\rm L})]^2$, where $\gamma_{\rm L}$ is the Lorentz factor which is approximately 1500 for Pb+Pb collisions at nucleon-nucleon c.m.s. energy $\sqrt{s_{\rm NN}}=2.76 \, {\rm TeV}$ and approximately 2700 for Pb+Pb collisions at nucleon-nucleon c.m.s. energy $\sqrt{s_{\rm NN}}=5.02 \, {\rm TeV}$. We model the form factor as the Fourier transform of the Woods-Saxon distribution~\cite{Woods:1954zz}, 
\begin{align}
F_A(t) & = \int d^3 r \rho_A (r) e^{i\boldsymbol{q}\cdot \boldsymbol{r}} \,, \label{eq:formfactor} \\
\rho_A (r) & = \frac{\rho_0}{1+e^{\frac{r-R_A}{d}}} \,,
\end{align}
taking $|\boldsymbol{q}| = \sqrt{|t|}$. We take $d=0.546\,\rm{fm}$~\cite{DeVries:1987atn} for the skin depth and for the nucleus radius $R_A$ we use the parametrization (see e.g. \cite{Helenius:2012wd}), 
\begin{equation}
R_A/\text{fm} = 1.12 \cdot A^{1/3} - 0.86 \cdot A^{-1/3}.
\end{equation}
The normalization $\rho_0$ is fixed by requiring that $F_A (0)=A$. 

When considering the $\gamma + p$ collisions we take the photoproduction cross section to be of the form \cite{Ivanov:2004vd},
\begin{align}
\sigma^{\gamma p \rightarrow Vp} (W) & = \frac{d\sigma^{\gamma p \rightarrow Vp}}{dt} \bigg|_{t=0} \int\limits_{0}^\infty dt' e^{-bt'} \label{eq:xsec4}
\end{align}
with \cite{Flett:2021xsl},  
\begin{equation}
    b \text{ GeV}^{2} = 4.9 + 4 \alpha_P' \ln \left( \frac{W}{W_0} \right) ,
\end{equation}
where $W_0 = 90$ GeV and $\alpha_P' = 0.06$. This parametrization grows more slowly with $W$ than that in Ref.~\cite{H1:2013okq}, but is still compatible with the HERA data for exclusive $J/\psi$ photoproduction. We have chosen the slope parameter $\alpha_P'$ to be compatible with Model 4 of~\cite{Khoze:2013dha} which fits a wider variety of elastic pp data. 

\subsection{Photoproduction Amplitude}
The NLO expressions for the matrix element $\mathcal{M}^{\gamma N \rightarrow VN}_A (W,t)$ for photoproduction are well established in the literature~\cite{Ivanov:2004vd,Jones:2015phd} and the more recent electroproduction results~\cite{Flett:2021xsl,Chen:2019uit,Flett:2021ghh} coincide with these in the limit of an on-shell photon. In these calculations the vector meson is considered as a composite particle of two heavy quarks in the non-relativistic approximation with zero relative velocity \cite{Berger:1980ni,Petrelli:1997ge,Bodwin:2002cfe,Braaten:2002fi}. The invariant matrix element can be written as 
\begin{equation} \label{Eq:FullAmplitude}
\begin{split}
        \mathcal{M}^{\gamma N \rightarrow VN}_A = &\frac{4\pi \sqrt{4\pi \alpha_{\rm QED}} e_Q (\varepsilon_V^* \cdot \varepsilon_\gamma )}{3\xi} \sqrt{\frac{\langle O_1 \rangle_V }{m_Q^3}}  I(\xi,t) \\
        =& \frac{C}{\xi} I(\xi,t) \,,
\end{split}
\end{equation}
where $\alpha_{\rm QED}$ is the fine-structure constant, $m_Q$ the mass of the heavy quark, $e_Q$ the fractional charge of the heavy quark, $\varepsilon_V$ the polarization vector of the produced vector meson, $\varepsilon_\gamma$ the polarization vector of the incoming photon, and $\langle O_1 \rangle_V$ is a non-relativistic QCD matrix element associated with the vector meson. Equation~\ref{Eq:FullAmplitude} defines the factor $C$ which we will use later. The value of $\langle O_1 \rangle_V$ is solved from the NLO expression for the vector-meson leptonic decay width \cite{Barbieri:1975ki,Bodwin:1994jh,Beneke:1997jm,Ivanov:2004vd}, 
\begin{equation}
    \Gamma (V \rightarrow l^+ l^-) = \frac{2e_Q^2 \pi \alpha_{\rm QED}^2}{3} \frac{\langle O_1 \rangle_V}{m_Q^2} \left[ 1 - \frac{8 \alpha_s(\mu_R)}{3\pi} \right]^2 \,,
\end{equation}
where $\alpha_s(\mu_R)$ is the QCD coupling at a renormalization scale $\mu_R$. The variable $\xi$ that appears in the Ji's parametrization of momenta \cite{Ji:1996nm}, is the so-called skewedness parameter. In the $t \ll M_V^2$ limit,
\begin{equation}
\xi = \frac{\zeta}{2-\zeta}, \text{ where } \zeta = \left( \frac{M_V}{W} \right)^2 \,.
\end{equation}
The function $I(\xi,t)$ is given by
\begin{equation}
\begin{split}
    I(\xi,t) = \int\limits_{-1}^1 dx [ &T_g (x,\xi) F^g (x,\xi,t,\mu_F) \\
    + &T_q (x,\xi ) F^{q,S} (x,\xi,t,\mu_F) ],
\end{split}
\end{equation}
where $T_g (x,\xi)$ and $T_q (x,\xi)$ are the hard-scattering coefficient functions corresponding to gluon and quark contributions \cite{Ivanov:2004vd},
\begin{equation}
    \begin{split}
        T_g (x,\xi) &= \frac{ \xi}{(x-\xi + i\epsilon ) (x+\xi -i\epsilon)}  \\
        &\times \Big[ \alpha_s (\mu_R) + \frac{\alpha_s^2 (\mu_R)}{4\pi} f_g \left( \frac{x-\xi +i\epsilon}{2\xi} \right) \Big] \,, \\
        T_q(x,\xi) &= \frac{2\alpha_s^2 (\mu_R)}{3\pi} f_q \left( \frac{x-\xi+i\epsilon}{2\xi} \right) \,.
    \end{split}
\end{equation}
Here the term proportional to $\alpha_s (\mu_R)$ in $T_g$ is the purely gluonic LO contribution and the rest in $T_g$ and the whole $T_q$ constitute the NLO contributions. The exact forms of the functions $f_g$ and $f_q$ are given in Refs.~\cite{Ivanov:2004vd,Jones:2015phd,Flett:2021xsl} and we will be using specifically those of Ref.~\cite{Ivanov:2004vd}. The parameter $\epsilon$ is positive and the function $I(\xi,t)$ is understood to be evaluated in the limit $\epsilon \rightarrow 0$. Finally, $F^{g} (x,\xi,t,\mu_F)$ is the gluon GPD and $F^{q,S} (x,\xi,t,\mu_F)$ is the quark singlet GPD given by
\begin{equation}
F^{q,S} (x,\xi,t,\mu_F) = \sum\limits_{q=u,d,s,c} F^q (x,\xi,t,\mu_F),
\end{equation}
where $\mu_F$ denotes the factorization scale. As we will consider factorization scales above the charm mass threshold, also the charm quarks are included in the above sum in conjunction with GPDs/PDFs defined in variable-flavour-number schemes. As indicated in Eq.~(\ref{eq:xsec1}), we will calculate the amplitude in the approximation in which $t=0$. In addition, in the current exploratory study we will approximate the GPDs by their values at $\xi=0$ so that we effectively replace the GPDs with PDFs,
\begin{align} 
F^{g} (x,0, 0,\mu_F) & = F^{g} (-x, 0 , 0,\mu_F) = xg(x,\mu_F) \,, \nonumber \\
F^q (x,0,0,\mu_F)    & = q(x,\mu_F)  \,, \label{Eq:GPDtoPDF} \\
F^q (-x,0,0,\mu_F) & = -\bar{q}(x,\mu_F) \,, \nonumber 
\end{align}
where $x \in [0,1]$, and $g(x,\mu_F)$ and $q(x,\mu_F)$ are the gluon and quark PDFs. The differential cross section can then be written as
\begin{equation}
\begin{split}
        \frac{d\sigma^{\gamma N \rightarrow VN}}{dt}\bigg|_{t=0} =&  \frac{\overline{|\mathcal{M}^{\gamma N \rightarrow VN}_A |^2}}{16\pi W^4} \\
        =& \frac{1}{W^4} \frac{4\pi^2 \alpha_{\rm QED} e_Q^2}{9 \xi^2} \left( \frac{\langle O_1 \rangle_V }{m_Q^3} \right) |I(\xi ,t=0)|^2,
\end{split}
\end{equation}
where
\begin{equation}\label{eq:I}
\begin{split}
        |I(\xi ,t=0)|^2 =& \Bigg| \int\limits_0^1 dx\Big[ 2 x g (x,\mu_F) T_g(x,\xi) \\
        +& T_q (x,\xi)  \sum\limits_q \left[ q(x,\mu_F) + \bar{q} (x,\mu_F) \Big]  \right] \Bigg|^2.
\end{split}
\end{equation}
In our analysis we take all constants, such as the mass and the decay width of the $J/\psi$, from the Particle-data-group listing \cite{Zyla:2020zbs}. The value of $\alpha_s (\mu_R)$ is taken from the LHAPDF interface~\cite{Buckley:2014ana} so that the coupling is taken consistently to be the same as the one used in defining the PDF values. The QED coupling, $\alpha_{\rm QED}$, is evaluated throughout the work up to one loop accuracy. In our framework, following Ref.~\cite{Ivanov:2004vd}, we explicitly set $M_V=2m_Q$ which is an inherent assumption in our non-relativistic approximation of the $J/\psi$ wavefunction. In our results for $|I|^2$ we consistently include both the real part and the imaginary part to the results. The integrals in Eq.~(\ref{eq:I}) are evaluated numerically by keeping the parameter $\epsilon$ finite but small enough so that the results are independent of its exact value. We have cross checked our numerical implementation against the method used in Ref.~\cite{Flett:2021xsl}. The factorization and renormalization scales are taken to be equal, $\mu = \mu_F=\mu_R$, and we consider scale variation between $\mu \in [m_Q,2m_Q]$. 

\subsection{Photon Flux}

The number of equivalent photons of energy $k$ at a fixed transverse distance $b$ from the center of a nucleus $A$ with $Z$ protons can be written as \cite{Vidovic:1992ik,Baltz:2007kq,Guzey:2013taa,Klein:2016yzr}
\begin{equation}
    N_{\gamma }^A (k,\Vec{b}) = \frac{Z^2\alpha_{\rm QED}}{\pi^2} \left| \int\limits_0^\infty d k_\perp \frac{k_\perp^2 F (k_\perp^2 + k^2/\gamma_L^2)}{k_\perp^2 + k^2/\gamma_L^2} J_1 (b k_\perp) \right|^2 ,
\end{equation}
where $F$ is the Fourier transform of the form factor in Eq.~(\ref{eq:formfactor}) normalized to one, $F(q)=F_A(q)/A$, and $J_1$ is the cylindrical modified Bessel function of the first kind. To obtain the minimum-bias flux appearing in the expression for the cross sections, e.g. in Eq.~(\ref{eq:xsec2}), we integrate over the entire impact-parameter plane multiplying $N_{\gamma }^A (k,\Vec{b})$ by the Glauber-type probability \cite{Florkowski:2010zz} of having no hadronic interaction, 
\begin{align}
    k \frac{dN_{\gamma }^A (k)}{dk} & = \int d^2 \Vec{b} N_{\gamma }^A (k,\Vec{b}) \Gamma_{AA}(\Vec{b}) \,, \label{eq:flux1} \\
    \Gamma_{AA}(\Vec{b}) & = \exp \left[ -\sigma_{\rm NN}(s) T_{AA} (\Vec{b}) \right] \,,
\end{align}
where $\sigma_{\rm NN}(s)$ is the total (elastic + inelastic) hadronic nucleon-nucleon cross section for which we use 90 (80) mb at $\sqrt{s_{\rm NN}} = 5.02\, (2.76)~{\rm TeV}$~\cite{Zyla:2020zbs}, and $T_{AA} (\Vec{b})$ is the nuclear overlap function
\begin{equation}
    T_{AA} (\Vec{b}) = \int d^2 \Vec{b}_1 T_A(\Vec{b}_1) T_A(\Vec{b} - \Vec{b}_1) \,,
\end{equation}
where $T_A (\Vec{b})$ is the nuclear thickness function, 
\begin{equation}
    T_A (\Vec{b}) = \int\limits_{-\infty}^\infty dz \rho_A (r) \,,
\end{equation}
with $r^2=z^2+\Vec{b}^2$ and $z$ being the longitudinal coordinate. The integrand in Eq.~(\ref{eq:flux1}) oscillates very rapidly at large values of $b$, and to improve the convergence we follow Ref.~\cite{Zha:2018ywo} by making use of the flux of photons from a point-like particle. In this case one takes the nuclear density to be a delta function, $\rho^{\rm pl}(\textbf{r}) = \delta^{3}(\textbf{r})$, which leads to \cite{vonWeizsacker:1934nji,Adeluyi:2012ph}, 
\begin{equation}
N^{\rm pl}_{\gamma / Z} (k,\Vec{b}) = \frac{Z^2 \alpha_{\rm QED}}{\pi^2} \frac{k^2}{\gamma_L^2} \left( K_1^2 (\zeta_R) + \frac{1}{\gamma_L^2} K_0^2 (\zeta_R) \right) \,, 
\end{equation}
where $K_0$ and $K_1$ are modified Bessel functions of the second kind, and 
\begin{equation}
    \zeta_R = \frac{kb}{\gamma_L} \,. 
\end{equation}
The integral over the impact-parameter plane with a condition $|{\vec b}| > b_{\rm min}$ is also well known \cite{Jackson:1998nia}, 
\begin{equation}
\begin{split}
k \frac{dN_{\gamma/Z}^{\rm pl} (k)}{dk}\Bigg|_{b_{\rm min}} & = \int_{b_{\rm min}}^\infty d^2 \Vec{b}  N^{\rm pl}_{\gamma / Z} (k,\Vec{b}) \\
& = \frac{2Z^2 \alpha_{\rm QED}}{\pi} \Big[ \zeta_R K_0 (\zeta_R) K_1 (\zeta_R) \\
&- \frac{\zeta_R^2}{2} (K_1^2 (\zeta_R ) -K_0^2 (\zeta_R)  ) \Big]_{b=b_{\rm min}} \,.
\end{split}
\end{equation}
We now rewrite Eq.~(\ref{eq:flux1}) by adding and subtracting the flux of photons from a point-like particle, 
\begin{equation}
\begin{split}
    k \frac{dN_{\gamma }^A (k)}{dk} & = \int d^2 \Vec{b} N_{\gamma }^A (k,\Vec{b}) \Gamma_{AA}(\Vec{b}) \\
    & + k \frac{dN_{\gamma/Z }^{\rm pl} (k)}{dk}\Bigg|_{b_{\rm min}} - k \frac{dN_{\gamma/Z }^{\rm pl} (k)}{dk}\Bigg|_{b_{\rm min}} \\
    & = k \frac{dN_{\gamma/Z }^{\rm pl} (k)}{dk}\Bigg|_{b_{\rm min}} +  \int\limits_0^{b_{\rm min}} d^2\Vec{b}~ N_{\gamma }^A (k,\Vec{b}) \Gamma_{AA}(\Vec{b}) \\
    & + \int\limits_{b_{\rm min}}^{\infty} d^2\Vec{b}~ \left[N_{\gamma }^A (k,\Vec{b})\Gamma_{AA}(\Vec{b}) - N^{\rm pl}_{\gamma / Z} (k,\Vec{b})\right] \,.
    \end{split}
\end{equation}
By taking $b_{\rm min} = 30\,{\rm fm}$ or higher, the last term will be negligible. Differences between this result and the point-like approximation have been studied e.g. in Refs.~\cite{Zha:2018ywo,Baltz:2007kq}.

\section{Results} \label{Sec:Results}

\subsection{Absolute magnitude and scale sensitivity of cross sections}

\begin{figure*}[ht]
    \centering
        \includegraphics[width=.8\textwidth]{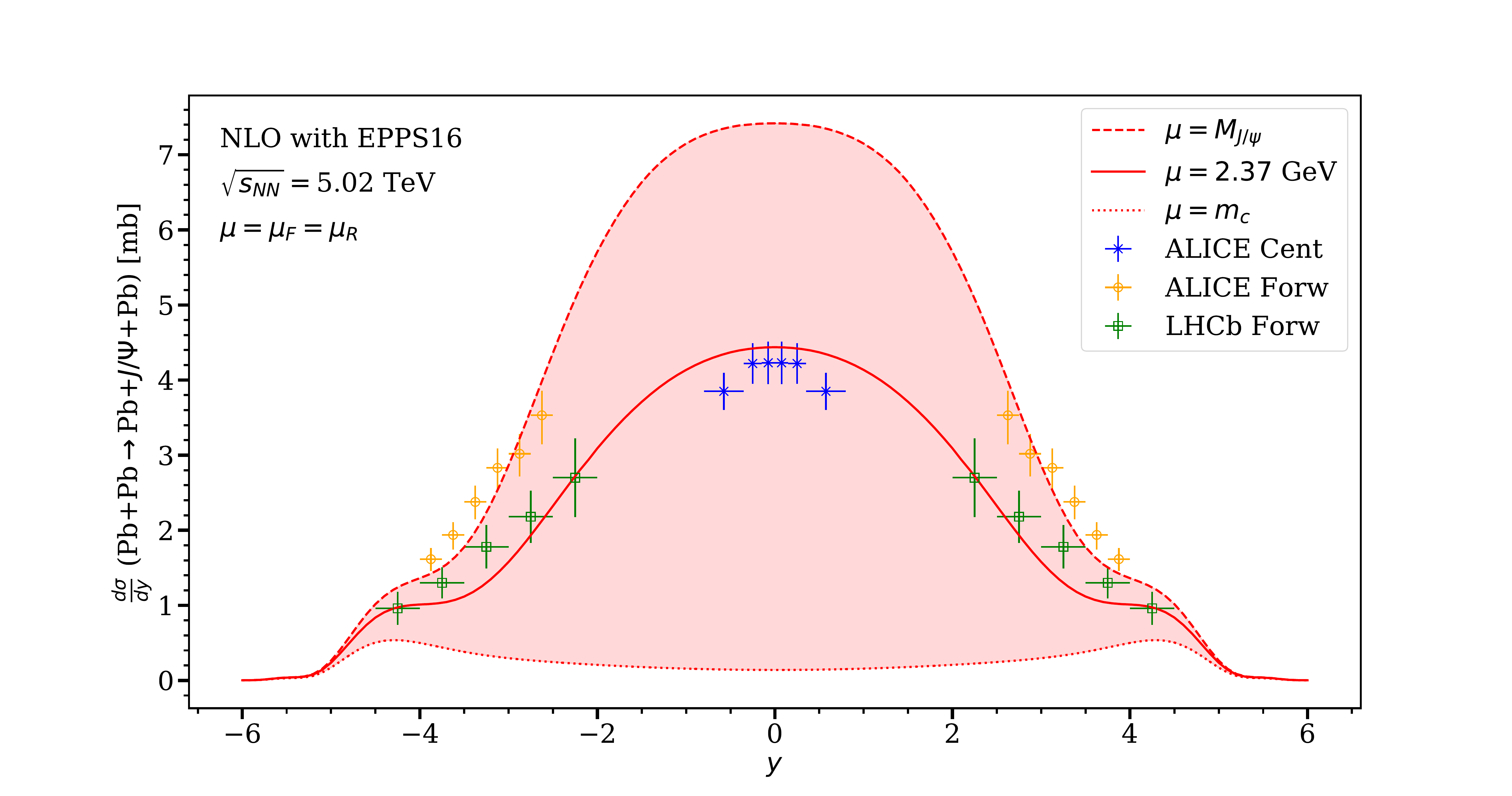}
        \includegraphics[width=.8\textwidth]{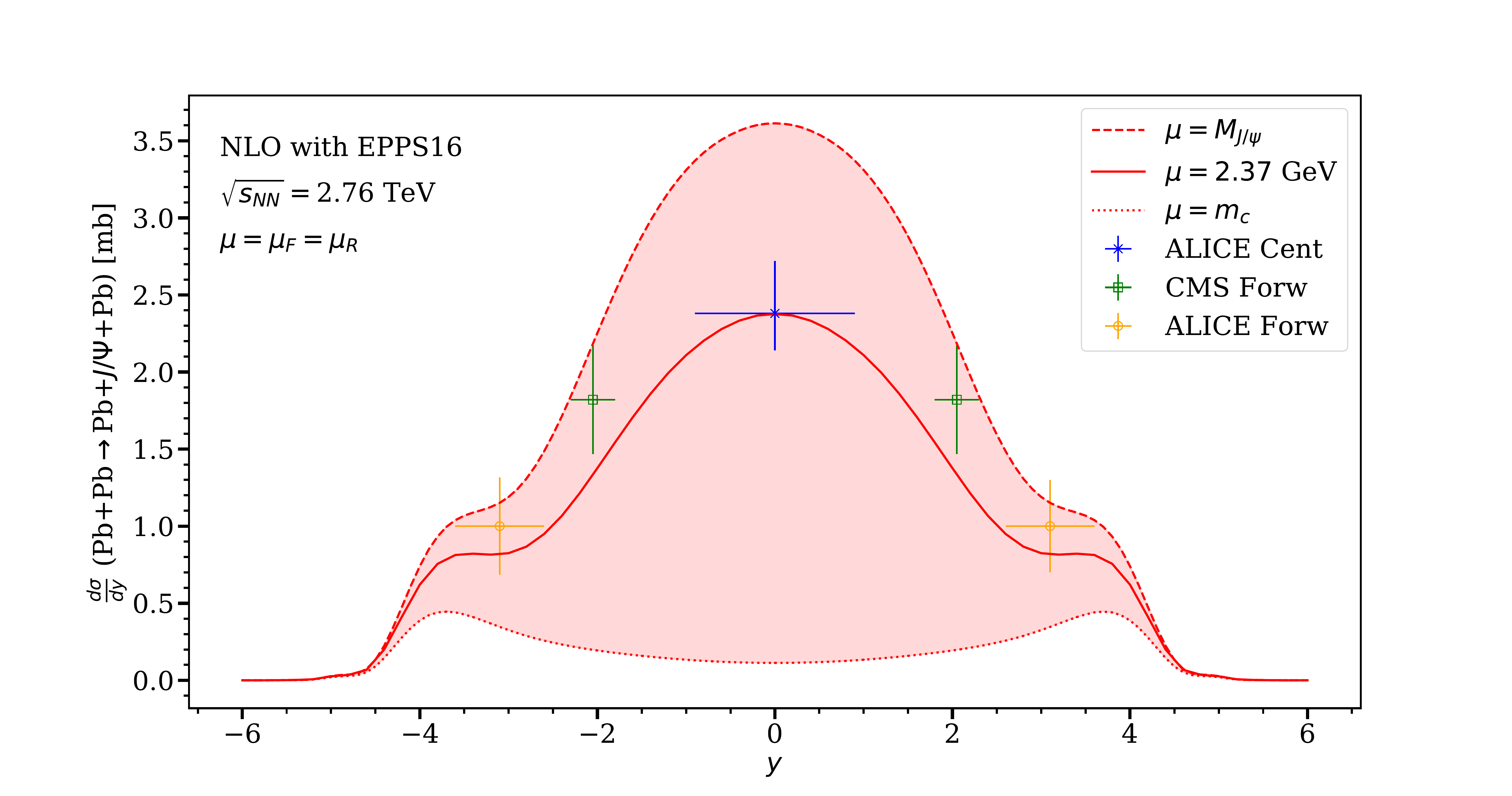}
    \caption{ \textbf{Upper panel:} The scale-choice uncertainty-envelope of the rapidity-differential exclusive $J/\psi$ photoproduction cross section in ultraperipheral Pb+Pb collisions at $\sqrt{s_{\rm NN}}= 5.02$ TeV, as a function of the $J/\psi$ rapidity $y$, calculated to NLO pQCD with the EPPS16 nPDFs \cite{Eskola:2016oht} and compared with the experimental data from Refs.~\cite{ALICE:2019tqa} (ALICE Forw), \cite{ALICE:2021gpt} (ALICE Cent) and \cite{LHCb:2021bfl} (LHCb Forw). The experimental data points are mirrored w.r.t. $y=0$, and their errorbars are obtained by adding the statistical and systematic errors in quadrature. The solid (red) curve shows the NLO result with our ``optimal" scale explained in the text.
    \textbf{Lower panel:} The same but at $\sqrt{s_{\rm NN}}= 2.76$ TeV and with experimental data from Refs.~\cite{ALICE:2012yye}  (ALICE Forw),  \cite{ALICE:2013wjo} (ALICE Cent) and \cite{CMS:2016itn} (CMS Cent). For the errorbars of the data, all given errors are added in quadrature.}
    \label{fig:Run1Run2Envelope}
\end{figure*}

First, we chart the uncertainty arising from the choice of the factorization/renormalization scale in the exclusive rapidity-differential $J/\psi$ photoproduction cross sections in ultraperipheral Pb+Pb collisions. Figure~\ref{fig:Run1Run2Envelope} shows the uncertainty envelopes that result from varying the scale $\mu = \mu_F = \mu_R$ from $M_{J/\psi}/2$ to  $M_{J/\psi}$ at $\sqrt{s_{\rm NN}}= 5.0$~TeV  (upper panel) and 2.76~TeV (lower panel), using the central set of the EPPS16 nPDFs \cite{Eskola:2016oht}. For comparison, the figure also shows the experimental LHC data measured at these energies at forward rapidities by ALICE \cite{ALICE:2019tqa,ALICE:2012yye}, LHCb \cite{LHCb:2021bfl} and CMS \cite{CMS:2016itn}, and at central rapidities by ALICE \cite{ALICE:2021gpt,ALICE:2013wjo}. The solid (red) lines in the middle-parts of the envelopes show the results with $\mu=0.76 M_{J/\psi} = 2.37$ GeV, a scale we have iteratively obtained by requiring a rough simultaneous fit to the data at both collision energies. In what follows, we call this ``optimal" scale, emphasizing however that its precise number bears no special significance but it depends e.g. on the assumed the GPD modeling details and nPDFs in general.

\begin{figure*}[ht]
    \centering
    \includegraphics[width=.8\textwidth]{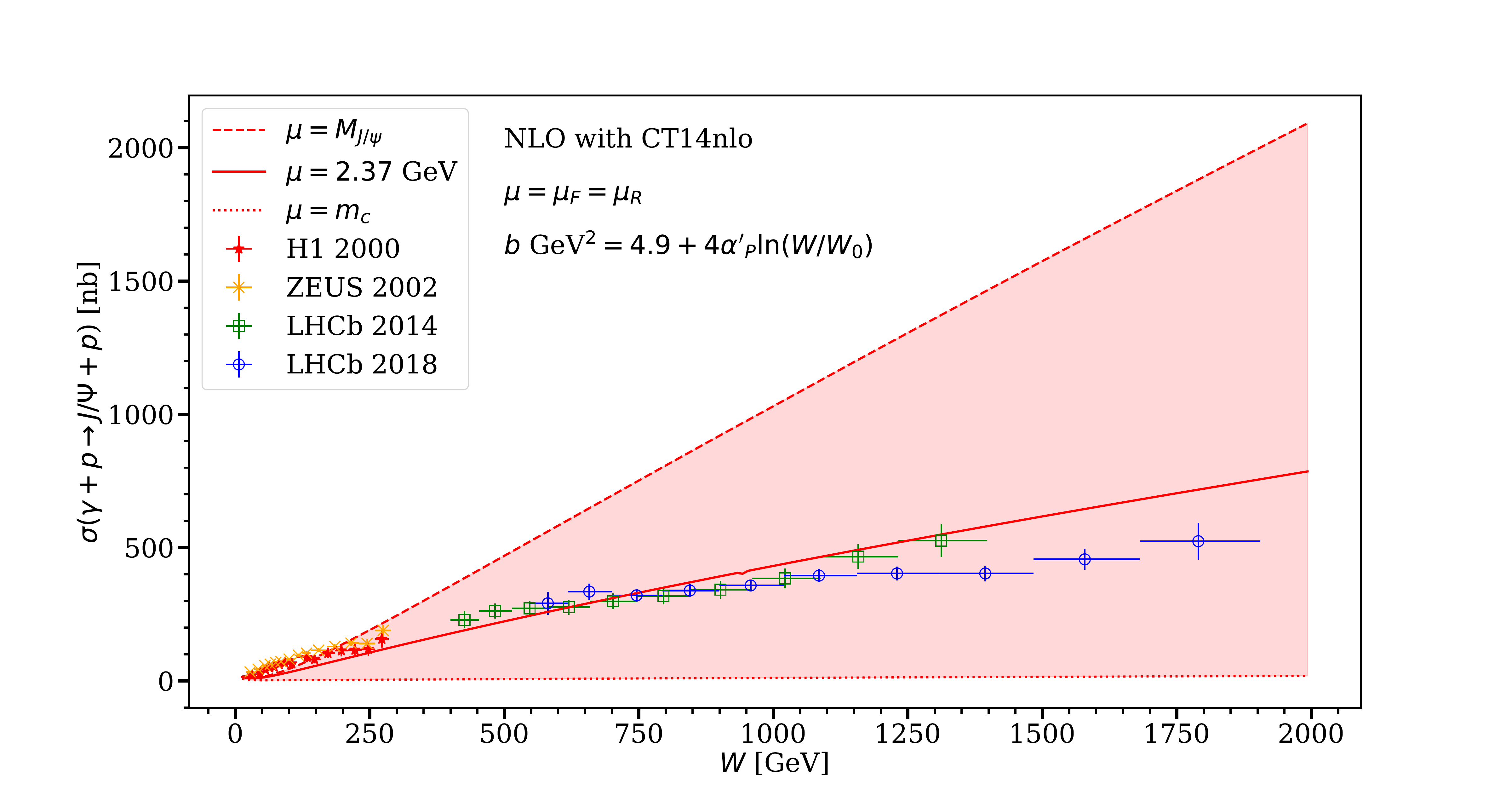}
\caption{The scale-choice uncertainty-envelope of exclusive $J/\psi$ photoproduction NLO cross sections in $ep$ and $pp$ collisions as a function of the photon-proton c.m.s. energy $W$, computed to NLO pQCD with the CT14NLO \cite{Dulat:2015mca} PDFs and compared against the experimental HERA data from 
 H1 \cite{H1:2000kis} and ZEUS~\cite{ZEUS:2002wfj}, and LHC data from LHCb~\cite{LHCb:2014acg,LHCb:2018rcm}. The solid (red) line corresponds to the ``optimal" scale explained in the text.}
    \label{fig:ep-baseline}
\end{figure*}

On the one hand,  as expected based on Ref.~\cite{Ivanov:2004vd}, we observe that the scale uncertainty remains quite large also here in the nuclear case. On the other hand then, it is interesting and quite encouraging that already with our current ``bare bones" GPD/PDF framework the NLO cross sections with entirely feasible scale choices $\mu={\cal O}(M_{J/\psi})$ not only are of the correct order of magnitude but actually some scale-choices can be found with which we can rather well reproduce the data at all rapidities and both collision energies. Earlier, especially with (ad hoc normalized) LO cross sections and the forward ALICE data at 5.02~TeV, this seemed not to be the case~\cite{ALICE:2021gpt}.   

Second, as a further check of our UPC results from the ``bare bones" GPD/PDF framework,  we study in Fig.~\ref{fig:ep-baseline} to what extent we can reproduce the exclusive $J/\psi$ photoproduction cross sections measured in $ep$ collisions at HERA and in $pp$ collisions at the LHC.\footnote{The photoproduction cross sections are extracted from the LHC $pp$ data through rather minimal modeling \cite{LHCb:2014acg,LHCb:2018rcm}.} The NLO cross sections here are, for consistency, computed with the CT14NLO PDFs \cite{Dulat:2015mca} which is the free-proton PDF set that the EPPS16 nPDFs are based on. The envelope shows again the uncertainty arising from varying the scale $\mu$ between $M_{J/\psi}/2$ and $M_{J/\psi}$. The HERA data in the figure are from H1~\cite{H1:2000kis} and ZEUS~\cite{ZEUS:2002wfj}, and the LHC data from LHCb~\cite{LHCb:2014acg,LHCb:2018rcm}. The solid (red) line in the middle of the envelope is again the NLO cross section computed with our ``optimal" scale which reproduced the nuclear data. 
As expected based on Ref.~\cite{Ivanov:2004vd} and other previous NLO studies of this process \cite{Jones:2015nna,Flett:2019nga,Flett:2019pux,Flett:2020duk,Flett:2021xsl}, the scale dependence is indeed large, and especially towards larger values of the photon-proton c.m.s. energy $W$ the data easily fall within the envelope. From the point of view of the nuclear UPCs the most relevant c.m.s.-energy region here is $W=10\dots700$ GeV (see the second $x$-axis in Fig.~\ref{fig:NLO_PlusMinDecomp} ahead). Interestingly, our framework with the ``optimal" scale leads to a rather reasonable overall agreement with the HERA/LHC $ep/pp$ data as well, except perhaps for the very lowest $W$ points. As suggested by earlier work \cite{Ivanov:2004vd,Jones:2015nna,Flett:2019nga,Flett:2019pux,Flett:2020duk,Flett:2021xsl}, there is room for GPD modeling testable against the $ep/pp$ data but given the large scale- and PDF-uncertainties (discussed in Fig.~\ref{fig:276-502-AllErrors} ahead), and also the exploratory nature of the present NLO study for UPCs of nuclei,  we leave this as future improvement. 
With Fig.~\ref{fig:ep-baseline}, it is also worth emphasizing that in the previous LO UPC studies one has typically normalized the LO cross sections to the HERA/LHC $ep/pp$ data and carried the obtained normalization factor then over to the UPC study, while in our current NLO study there are \textit{no} ad hoc normalization factors.

\begin{figure*}[ht]
    \centering
     \includegraphics[width=.8\textwidth]{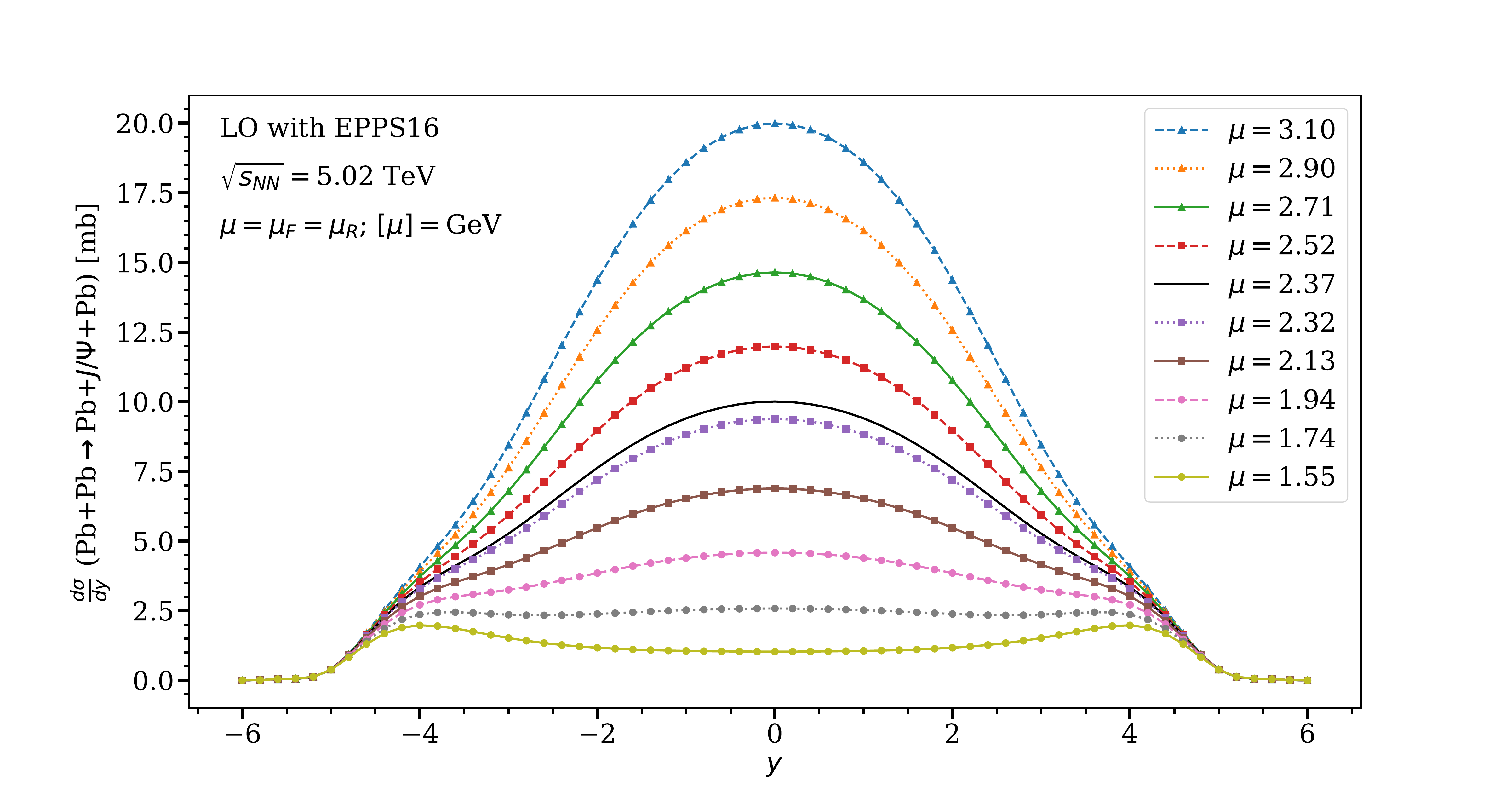}
     \includegraphics[width=.8\textwidth]{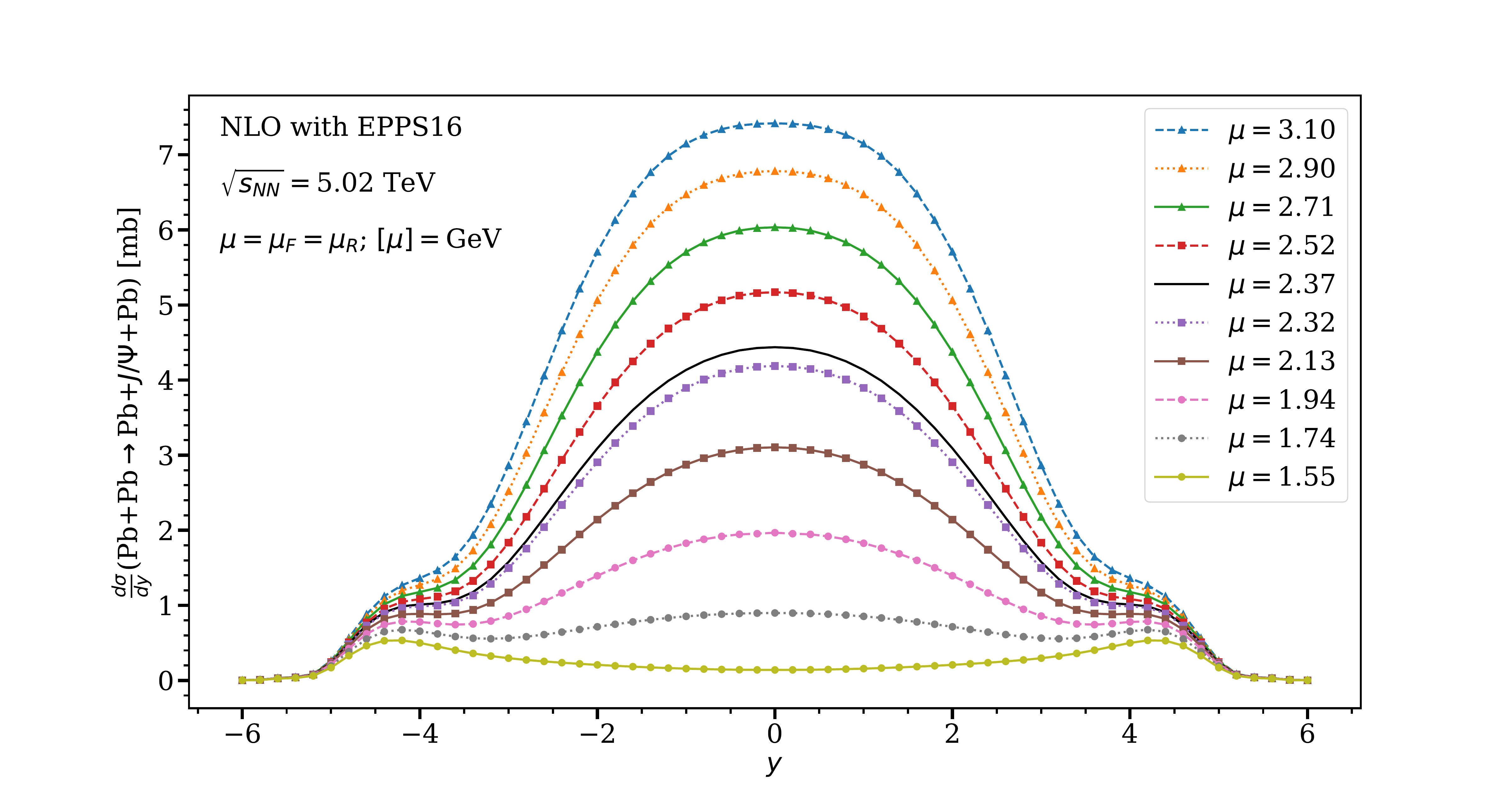}
      \caption{\textbf{Upper panel:} Rapidity-differential exclusive $J/\psi$ photoproduction cross sections in Pb+Pb UPCs at $\sqrt{s_{\rm NN}}=5.02$~TeV, as a function of the rapidity $y$, computed at LO pQCD with the EPPS16 nPDFs at various fixed scales $\mu$. The lowest- and highest-scale results here give the envelope shown in Fig.~\ref{fig:Run1Run2Envelope}. The result with our ``optimal" scale is shown by the solid curve. \textbf{Lower panel:} The same but at NLO pQCD.} 
    \label{fig:LOandNLOfan}
\end{figure*}

Third, we investigate the stability of the rapidity-differential $J/\psi$ photoproduction cross sections in Pb+Pb UPCs, i.e. the changes in the magnitude and shape, and in the scale-dependence of the cross sections, when moving from LO to NLO in pQCD.  These questions are answered by Fig.~\ref{fig:LOandNLOfan}, where we show the rapidity-differential cross sections computed with various fixed scales $\mu$ between $M_{J/\psi}/2$ and $M_{J/\psi}$ in the LO and NLO cases (upper and lower panels, correspondingly). To be exact, the LO here refers to the purely gluonic Born-term contribution which enters the full NLO result. For the  computation, we again use the EPPS16 nPDFs.
We observe that the overall effect of the NLO terms is to reduce the LO cross sections rather significantly, at the ``optimal" scale by a factor of 2.3 at mid-rapidity, and by a factor of 3.3 at $y=\pm 4$. We also see that the studied scale-variation causes about a factor of 20 change in the LO case while in the full NLO result the change is about a factor of 50. These results confirm the expectations based on Ref.~\cite{Ivanov:2004vd} also now in the nuclear UPC case, that at the low scales of $\mu={\cal O}(M_{J/\psi})$ the NLO contributions do not stabilize the results, yet, but bring the cross sections nevertheless into the right direction.
Interestingly, as seen in Fig.~\ref{fig:LOandNLOfan}, also the whole shape of both the LO and the NLO results is quite sensitive to the scale $\mu$, and again perhaps even more so at NLO, in this scale range. In the LO case, the strong scale dependence can be traced back mainly to the rapidly changing gluon distributions, while in the NLO terms the scale $\mu$ resides both in the pQCD matrix elements and in the PDFs. In particular, as we will soon see, in the NLO cross sections the rapidly evolving small-$x$ \textit{quark} PDFs start to play a surprisingly important, and at mid-rapidities even a dominant, role. 

\begin{figure*}[t]
    \centering
     \includegraphics[width=.8\textwidth]{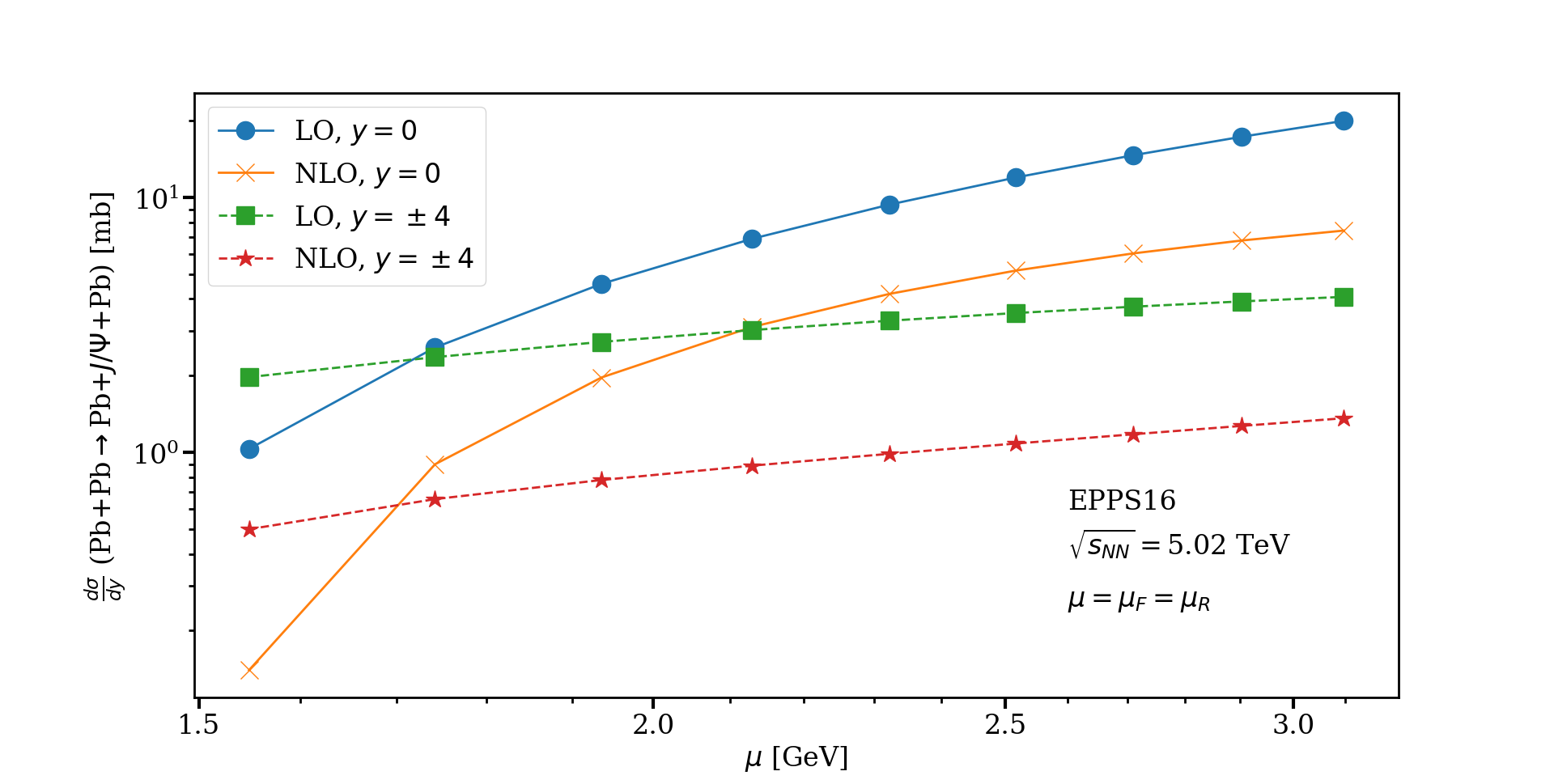}
      \caption{The NLO (crosses and stars) and LO (filled circles and boxes) rapidity-differential cross sections of Fig.~\ref{fig:LOandNLOfan} at $y=0$ (solid lines) and $y=\pm4$ (dashed lines), as a function of the scale choice $\mu$.			} 
    \label{fig:LOandNLOscaledependence}
\end{figure*}

To analyse the scale dependence of our LO and full NLO results and their interrelation further, we plot in Fig.~\ref{fig:LOandNLOscaledependence} the computed rapidity-differential cross sections at fixed rapidities $y=0$ and at $y=\pm 4$ as a function of the scale $\mu$. As we see, at $y=0$ the scale dependence at low scales is stronger in the NLO than in the LO results but towards higher scales it actually becomes weaker. At $y=\pm4$ we see the scale dependence being stronger in NLO at all scales studied. Thus, whether the scale dependence is improved (tamed) when going from LO to NLO depends on the rapidity $y$ and potentially also the scale-choice region. Another interesting observation is that our ``optimal" scale $\mu=2.37$ GeV is right in the region where the scale dependence at $y=0$ turns from stronger to weaker relative to LO, i.e. where the LO and NLO results are closest to each other. At $y=\pm4$, however, we do not find a similar taming effect to take place. This figure also shows how the NLO/LO ratio (``K-factor") is not a constant as a function of the scale, and certainly not a constant as a function of the $J/\psi$ rapidity.

\subsection{Complex structure of the cross section}

\begin{figure*}[ht]
    \centering
    \includegraphics[width=.8\textwidth]{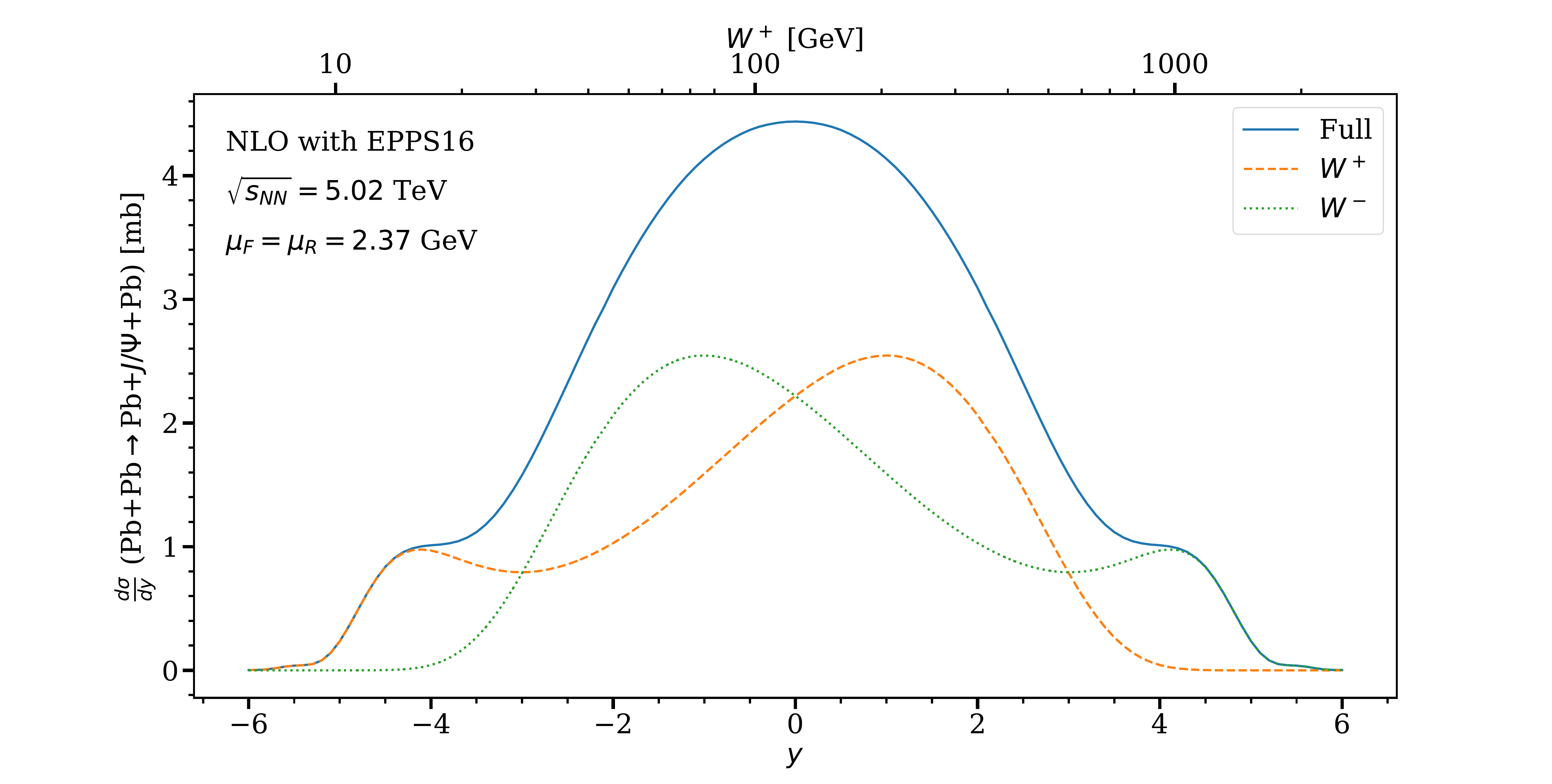}
\caption{Contributions from the $W^+$ (dashed, red curve) and $W^-$ (dotted, green curve)	terms in Eq.~(\ref{XS_plus_minus}) to the NLO exclusive rapidity-differential $J/\psi$ photoproduction cross section in 5.02 TeV Pb+Pb UPCs as a function of the $J/\psi$ rapidity $y$, computed using EPPS16 nPDFs and with our ``optimal" scale.	The second $x$ axis on the top shows the values of $W^+$ corresponding to each $y$.} \label{fig:NLO_PlusMinDecomp}
\end{figure*}

Next, we discuss the very interesting consequences of the complex structure of the rapidity-differential $J/\psi$ photoproduction cross sections in 5.02~TeV Pb+Pb UPCs. First, in Fig.~\ref{fig:NLO_PlusMinDecomp} we study the $k^\pm$  contributions in Eq.~(\ref{XS_plus_minus}) to the rapidity-differential $J/\psi$ photoproduction NLO cross section in 5.02 TeV Pb+Pb UPCs, computed with EPPS16 at our ``optimal" scale. The photon-proton c.m.s. energy corresponding to the photon energies $k^\pm$ in Eq.~(\ref{XS_plus_minus}) are denoted by $W^\pm$ in what follows. As indicated by the second $x$-axis at the top of Fig.~\ref{fig:NLO_PlusMinDecomp}, $W^+$ ($W^-$) increases to the right (left). As we saw in Fig.~\ref{fig:ep-baseline} above, the photoproduction cross section in the $k^\pm$  terms of Eq.~(\ref{XS_plus_minus}) increases as a function of $W^\pm$, correspondingly. The photon flux, however, decreases rapidly as a function of the energy $W^\pm$ (see e.g. Fig.~3 of \cite{Guzey:2013taa}), causing the nonmonotonous behaviour of the two symmetric contributions as is seen in Fig.~\ref{fig:NLO_PlusMinDecomp}. Looking at the $W^+$ curve (dashed, red) we see that first at backward-most rapidities the photon flux is high enough to produce a noticeable cross section in spite of the smallness of the photoproduction cross section there. Also the $t$ integral of the squared form factor of Eq.~(\ref{eq:formfactor}) reaches non-negligible values by $y\sim -4$, which also contributes to the initial rise of the cross section at backward-most rapidities. Then in the ``shoulder" region the decrease of the photon flux wins over the increase of the photoproduction cross section, causing the small dip seen in the figure. Approaching then mid-rapidities, the increase of the photoproduction cross section now wins over the decrease of the photon flux, until eventually towards forward-most rapidities the photon flux decrease again dominates and the resulting cross section dies out. For the $W^-$ component (dotted green curve), the behaviour is a mirror image of this, and the final result (solid blue curve) is a combination of the $W^\pm$ contributions as seen in the figure.

\begin{figure*}[ht]
\centering
 \includegraphics[width=.8\textwidth]{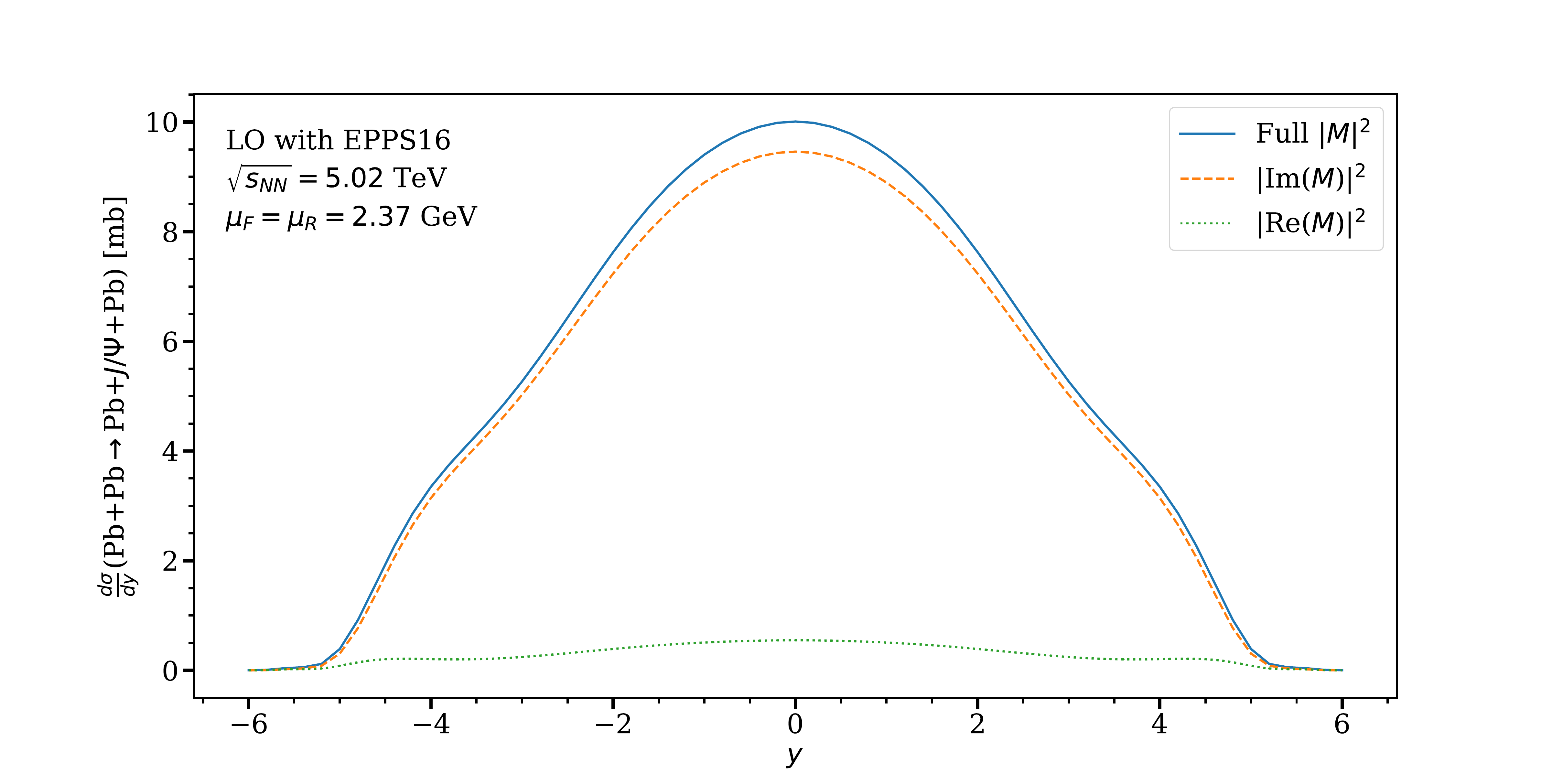}
 \includegraphics[width=.8\textwidth]{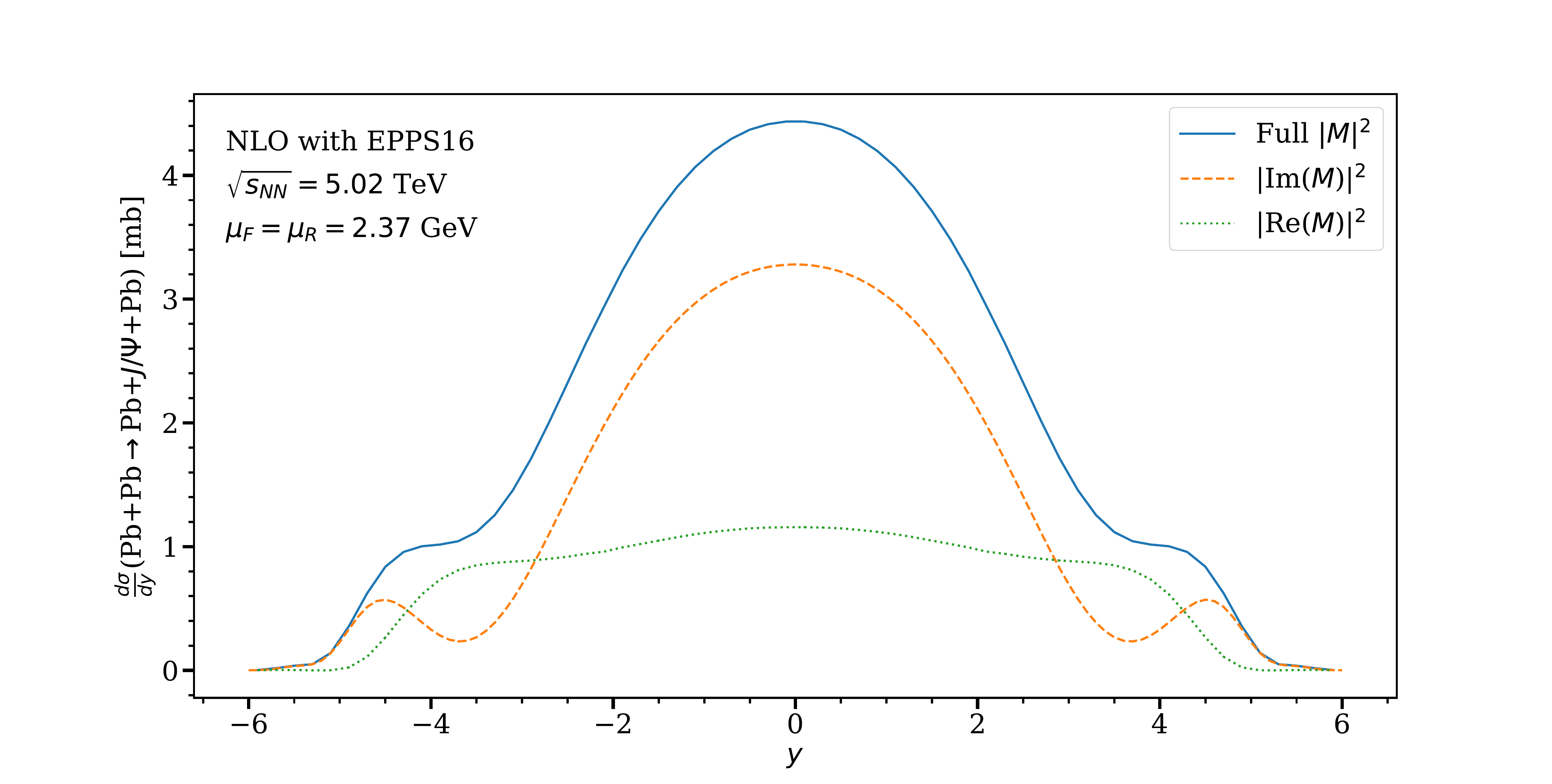}
\caption{\textbf{Upper panel:} Contributions from the real part (dotted green curve) and imaginary part (dashed red curve) of the amplitude to the LO exclusive rapidity-differential $J/\psi$ photoproduction cross section in 5.02 TeV Pb+Pb UPCs (solid blue curve) as a function of the rapidity, computed using the EPPS16 nPDFs at our ``optimal" scale.
\textbf{Lower panel:} The same but in NLO.}
    \label{fig:LOandNLOReIm}
\end{figure*}

Second, we quantify the contributions from the imaginary and real parts of the amplitude. The decomposition of the full result ($\propto |{\cal M}|^2$)  into the contributions from the real part ($\propto |\text{Re}({\cal M})|^2$) and the imaginary part ($\propto |\text{Im}({\cal M})|^2$) for both the LO and NLO cross sections is shown in Fig.~\ref{fig:LOandNLOReIm}. These results are again obtained with the EPPS16 nPDFs and fixing $\mu$ to our ``optimal" scale. The LO here again refers to the Born term contributions entering the full NLO result. As the upper panel shows, in the LO case where only gluons contribute, we confirm -- at least for gluon PDFs of a modest small-$x$ rise, such as those in EPPS16/CT14NLO -- the general claim that the contribution from the imaginary part of the amplitude clearly dominates at all rapidities. However, as the lower panel shows, the situation changes rather dramatically for the NLO cross sections: At mid-rapidity the contribution from the real part of the amplitude is about a quarter, which clearly is not anymore negligible. Towards forward/backward rapidities the real-part contributions become even more important and, as seen in the figure, there is a region at large/small rapidities where they dominate over the imaginary-part contributions. These findings are also consistent with those of Ref.~\cite{Ivanov:2004vd}, see Fig.~17 there. The message from Fig.~\ref{fig:LOandNLOReIm} is clear: both the imaginary and real parts of the amplitude must be accounted for in the calculation of these cross sections. 

\begin{figure*}[ht]
    \centering
    \includegraphics[width=.8\textwidth]{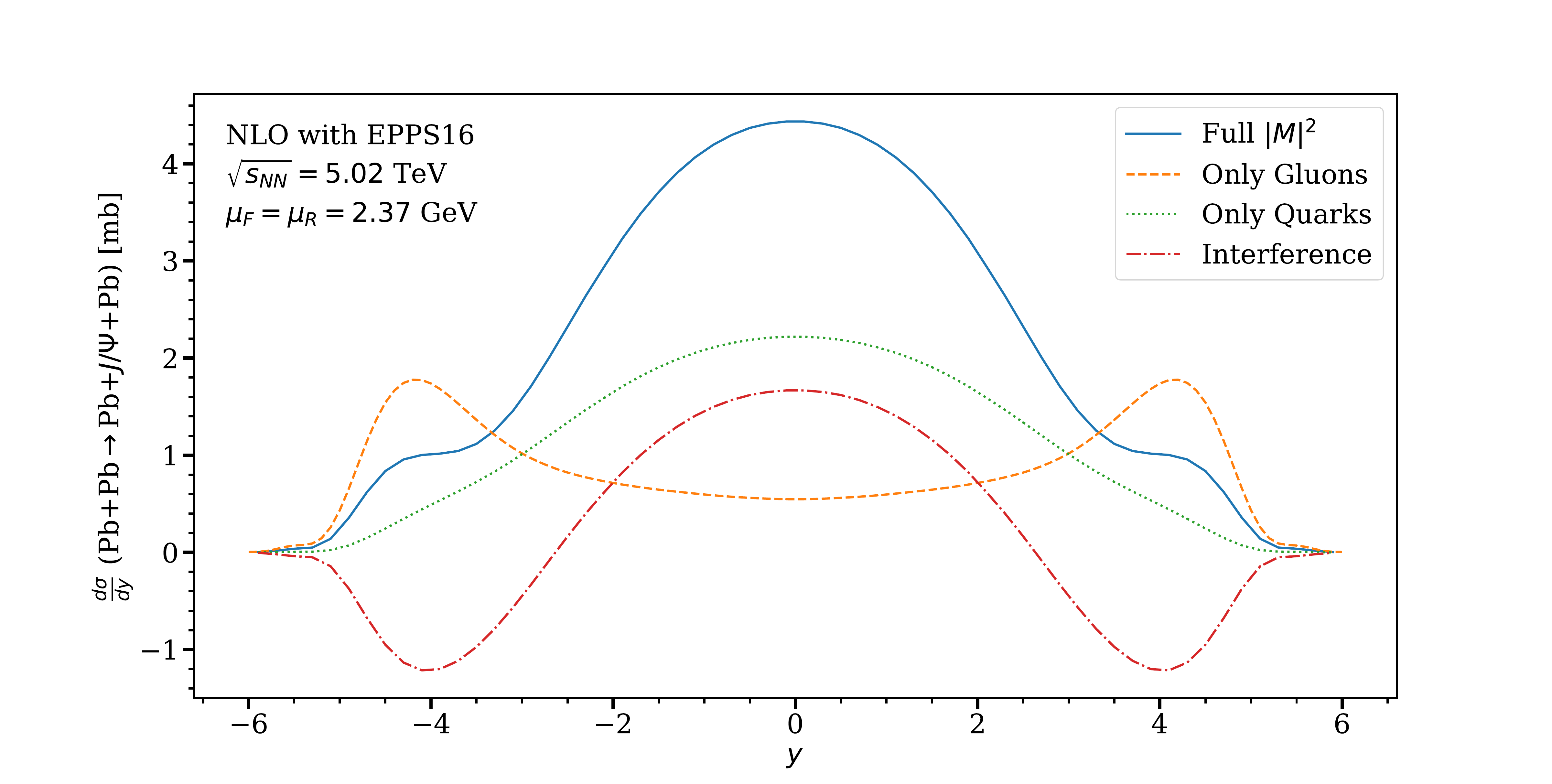}
    \caption{Decomposition of the exclusive  rapidity-differential $J/\psi$ photoproduction cross section, computed with EPPS16 nPDFs at our ``optimal" scale, in 5.02 TeV Pb+Pb UPCs (solid blue curve ``Full $|{\cal M}|^2$") into the contributions with zero quark distributions (dashed orange curve ``Only Gluons"), with zero gluon distributions (dotted green curve ``Only Quarks") and the one with a mixing of the quark and gluon distributions in the square of the full NLO amplitude (red dashed-dotted curve ``Interference"). }
    \label{fig:NLO_EPPS_oGoQ}
\end{figure*}

Third, in Fig.~\ref{fig:NLO_EPPS_oGoQ}, we investigate the breakdown of the computed $J/\psi$ photoproduction NLO cross section in 5.02 TeV Pb+Pb UPCs into the quark and gluon contributions, using EPPS16 and our ``optimal" scale. The solid (blue) curve labelled ``Full $|{\cal M}|^2$" is the full NLO cross section of Fig.~\ref{fig:LOandNLOReIm}, while the dashed red (dotted green) curve labelled ``Only Gluons" (``Only Quarks") is obtained by setting the quark (gluon) distributions to zero. The dashed-dotted curve labelled ``Interference" corresponds to the remaining contribution from the cross section pieces that contain both quarks and gluons. As shown by Fig.~\ref{fig:NLO_EPPS_oGoQ}, at mid-rapidity the quarks-only contribution dominates over the gluons-only by a factor of four (!), and the quark-gluon term over the gluons-only by a factor of three. Towards forward/backward-most rapidities the gluons-only contributions become the dominant ones, and we can see that the gluon-quark term also changes its sign when going from mid-rapidity to forward/backward rapidities. 

Recalling the original attraction of the exclusive $J/\psi$ photoproduction in electron-proton collisions and in nuclear UPCs as an exceptionally efficient probe of small-$x$ gluon distributions, the results in Fig.~\ref{fig:NLO_EPPS_oGoQ}  appear at first sight somewhat surprising. Especially the quark dominance at mid-rapidity seems to be in a direct contradiction with the original LO-based gluon-probe suggestion, and in fact also with our expectation that small-$x$ gluons \textit{should} after all dominate also the NLO contributions! 
  
\begin{figure*}[ht]
    \centering
    \includegraphics[width=.8\textwidth]{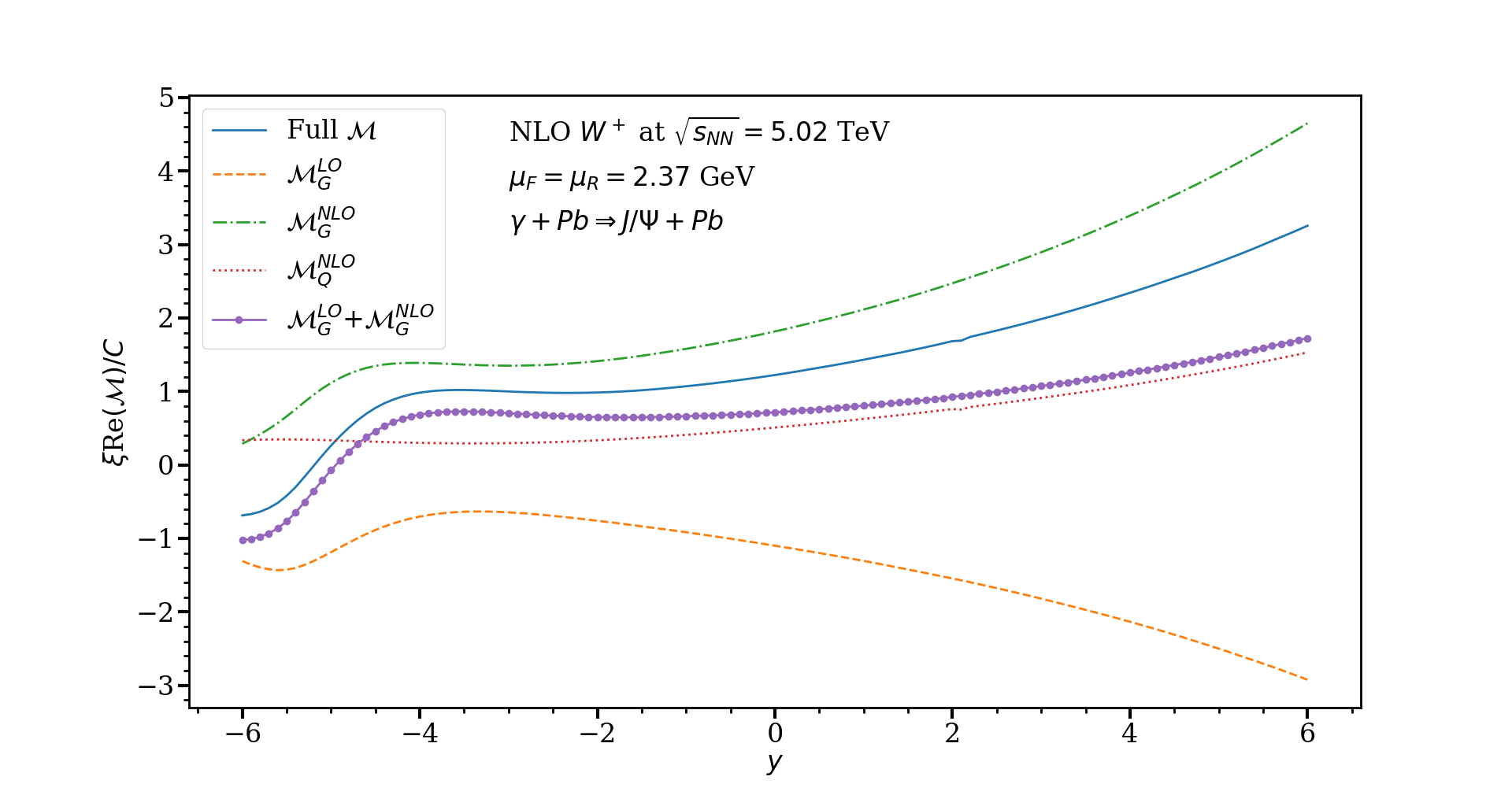}
    \includegraphics[width=.8\textwidth]{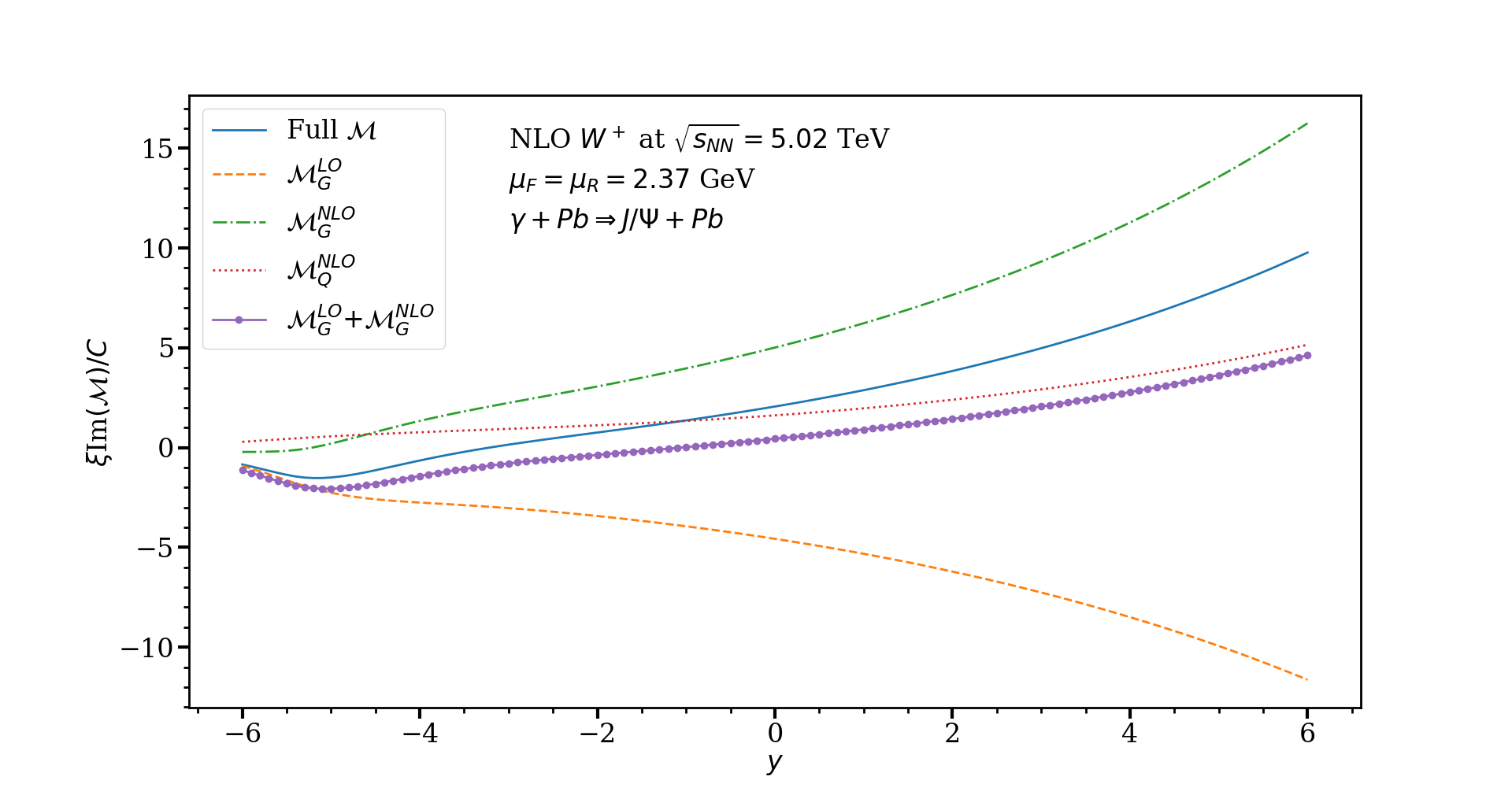}
    \caption{\textbf{Upper panel}: The $\xi/C$-scaled real parts of the full amplitude $\mathcal{M}$ (solid blue curve), LO gluon term $\mathcal{M}_G^\text{LO}$ (dashed orange), NLO gluon term $\mathcal{M}_G^\text{NLO}$ (dashed dotted green), sum of the LO and NLO gluon terms $\mathcal{M}_G^\text{LO}+ \mathcal{M}_G^\text{NLO}$ (solid purple with filled circles), and NLO quark term $\mathcal{M}_Q^\text{NLO}$ (dotted red), as a function of the $J/\psi$ rapidity $y$, for the contribution $W^+$. For the definition of the scaling factor, see Eq.~(\ref{Eq:FullAmplitude}).
\textbf{Lower panel:} The same but for the imaginary parts of the amplitudes. Notice the different vertical scale.}
    \label{fig:Re_Im_EPPS16}
\end{figure*}

A better understanding of this clearly calls for a more detailed look at the individual contributions in the LO and NLO amplitudes. For this purpose, we write the full NLO amplitude in terms of the LO gluon part $\mathcal{M}_G^{\text{LO}}$ and NLO gluon and quark parts $\mathcal{M}_G^{\text{NLO}}$ and $\mathcal{M}_Q^{\text{NLO}}$, 
\begin{equation}
\mathcal{M} = 
\mathcal{M}_G^{\text{LO}} + \mathcal{M}_G^{\text{NLO}} + \mathcal{M}_Q^{\text{NLO}},
\end{equation}
so that the squared amplitude entering the cross section becomes
\begin{equation}
\begin{split}
|\mathcal{M}|^2 &= 
|\mathcal{M}_G^{\text{LO}} + \mathcal{M}_G^{\text{NLO}}|^2 + |\mathcal{M}_Q^\text{NLO}|^2 \\
& + 2\Big[\text{Re}(\mathcal{M}_G^{\text{LO}} + \mathcal{M}_G^\text{NLO})\text{Re}(\mathcal{M}_Q^\text{NLO}) \\
&+  \text{Im}(\mathcal{M}_G^{\text{LO}} + \mathcal{M}_G^\text{NLO})\text{Im}(\mathcal{M}_Q^\text{NLO})\Big].
\label{Im_and_Re_parts}
\end{split}
\end{equation}
The gluons-only contribution in Fig.~\ref{fig:NLO_EPPS_oGoQ} comes from the term 
\begin{equation}
\begin{split}
|\mathcal{M}_G^\text{LO} + \mathcal{M}_G^\text{NLO}|^2 
&=  [\text{Re}(\mathcal{M}_G^\text{LO}) + \text{Re}(\mathcal{M}_G^\text{NLO})]^2 \\
 &+ [\text{Im}(\mathcal{M}_G^\text{NLO}) + \text{Im}(\mathcal{M}_G^\text{NLO})]^2
 \end{split}
\end{equation}
and the quarks-only contribution from 
\begin{equation}
|\mathcal{M}_Q^\text{NLO}|^2 
=  [\text{Re}(\mathcal{M}_Q^{NLO})]^2 + [\text{Im}(\mathcal{M}_Q^{NLO})]^2,
\end{equation}
while the gluon-quark interference contribution corresponds to the third term on the r.h.s. of  Eq.~(\ref{Im_and_Re_parts}).

Figure~\ref{fig:Re_Im_EPPS16} shows the above real and imaginary parts of the amplitude, multiplied with the factor $\xi/C$ (see Eq.~\ref{Eq:FullAmplitude}), as a function of the rapidity corresponding to $W^+$.\footnote{Recall that the photon flux and form factor do not enter here.} This figure finally reveals exactly what is behind the quark and gluon contributions in Fig.~\ref{fig:NLO_EPPS_oGoQ}: 
In their \textit{absolute} values, the LO and NLO gluon amplitudes  $\mathcal{M}_G^\text{LO}$ and $\mathcal{M}_G^\text{NLO}$
indeed \textit{do} clearly dominate over the quark contribution $\mathcal{M}_Q^\text{NLO}$ both in the real and imaginary parts. However, due to their opposite signs, the LO and NLO gluon amplitudes \textit{cancel} to a large degree both in the real and imaginary parts. The exact efficiency of the cancellation depends on the rapidity ($W^+$), and  $\text{Im}(\mathcal{M}_G^\text{LO}) + \text{Im}(\mathcal{M}_G^\text{NLO})$
changes its sign from plus to minus when approaching backward rapidities, which causes the sign change of the quark-gluon mixing term in Fig.~\ref{fig:NLO_EPPS_oGoQ}.
Let us look at the following three example-rapidities:

\textbullet~ At $y=0$, where the $W^\pm$ components contribute equally (see Fig.~\ref{fig:NLO_PlusMinDecomp}), the cancellation of the gluon terms is coincidentally (that is, with these PDFs) almost perfect in the imaginary part, so that 
\begin{equation}
\begin{split}
 [\text{Im}(\mathcal{M}_G^\text{LO} +\mathcal{M}_G^\text{NLO})]^2
&\ll [\text{Re}(\mathcal{M}_G^\text{LO} +\mathcal{M}_G^\text{NLO})]^2 \\
& \lesssim [\text{Re}(\mathcal{M}_Q^\text{NLO})]^2 \\
&\ll [\text{Im}(\mathcal{M}_Q^\text{NLO})]^2,
\end{split}
\end{equation}
which makes the imaginary part of the quark amplitude dominate the cross section in Fig.~\ref{fig:NLO_EPPS_oGoQ}. In the quark-gluon mixing term then the product of the imaginary parts dominates over the product of the real parts, and due to the large 
$\text{Im}(\mathcal{M}_Q^\text{NLO})$ the quark-gluon contribution dominates over the gluons-only term. 

\textbullet~ At $y\approx-3$, Fig.~\ref{fig:NLO_PlusMinDecomp} indicates that the $W^\pm$ contributions are equally important, so that Fig.~\ref{fig:Re_Im_EPPS16} should be read both at $y\approx -3$ and $y\approx +3$. The 
squared amplitude is larger for $y=3$ but the rapid decrease of the $W^-$-component's photon flux and nuclear form factor towards negative rapidities now suppresses the $W^-$ component so that it becomes of the same magnitude as the $W^+$ component whose squared amplitude is smaller but photon flux correspondingly larger. As seen in Figs.~\ref{fig:LOandNLOReIm} and \ref{fig:NLO_EPPS_oGoQ}, 
as a result of these competing effects the real and imaginary parts of the amplitude, as well as quarks and gluons, then contribute equally to the rapidity-differential cross section at $y\approx-3$.

\textbullet~ At $y\approx-4$,  where the cross section is dominated by the $W^+$ component as seen in Fig.~\ref{fig:NLO_PlusMinDecomp}, the LO and NLO gluon terms cancel to a much smaller degree both in the real and imaginary parts, and the hierarchy becomes 
\begin{equation}
\begin{split}
[\text{Re}(\mathcal{M}_Q^\text{NLO})]^2 
&\ll 
[\text{Re}(\mathcal{M}_G^\text{LO} + \mathcal{M}_G^\text{NLO})]^2 \\
&\lesssim 
[\text{Im}(\mathcal{M}_Q^\text{NLO})]^2 \\
&\ll 
[\text{Im}(\mathcal{M}_G^\text{LO} +\mathcal{M}_G^\text{NLO})]^2,
\end{split}
\end{equation}
causing the gluons-only terms to dominate over the quarks-only by a factor of four. In this case, the sizable quark-gluon mixing term is deeply negative because of the large negative term   
$\text{Im}(\mathcal{M}_G^\text{LO}) + \text{Im}(\mathcal{M}_G^\text{NLO})$. It is again the negative sign of this term that in the full amplitude causes  
$[\text{Re}(\mathcal{M})]^2 \gtrsim [\text{Im}(\mathcal{M})]^2$, seen in Fig.~\ref{fig:Re_Im_EPPS16} and in the lower panel of Fig.~\ref{fig:LOandNLOReIm} at $y=-4...-3$.

As shown by Figs.~\ref{fig:NLO_PlusMinDecomp}-\ref{fig:Re_Im_EPPS16},
the full NLO cross section thus has a very detailed complex structure with interplays between the photoproduction cross section, the photon flux and the nuclear form factor, between the $W^\pm$ components, and especially between the various contributions from the real and imaginary parts of the amplitude. The key to understand the obtained rapidity-differential cross sections is the degree of cancellation of the LO and NLO gluon contributions of opposite signs. We have also checked that the situation is qualitatively the same for the 2.76 TeV collision energy, and that the real part contributions become slightly more important for all values of $y$ than for the 5.02 TeV case. We have also checked that in the case of no nuclear effects, the situation remains qualitatively the same.

\subsection{Nuclear effects and PDF uncertainties in the cross section}

\begin{figure*}[ht]
    \centering
    \includegraphics[width=.8\textwidth]{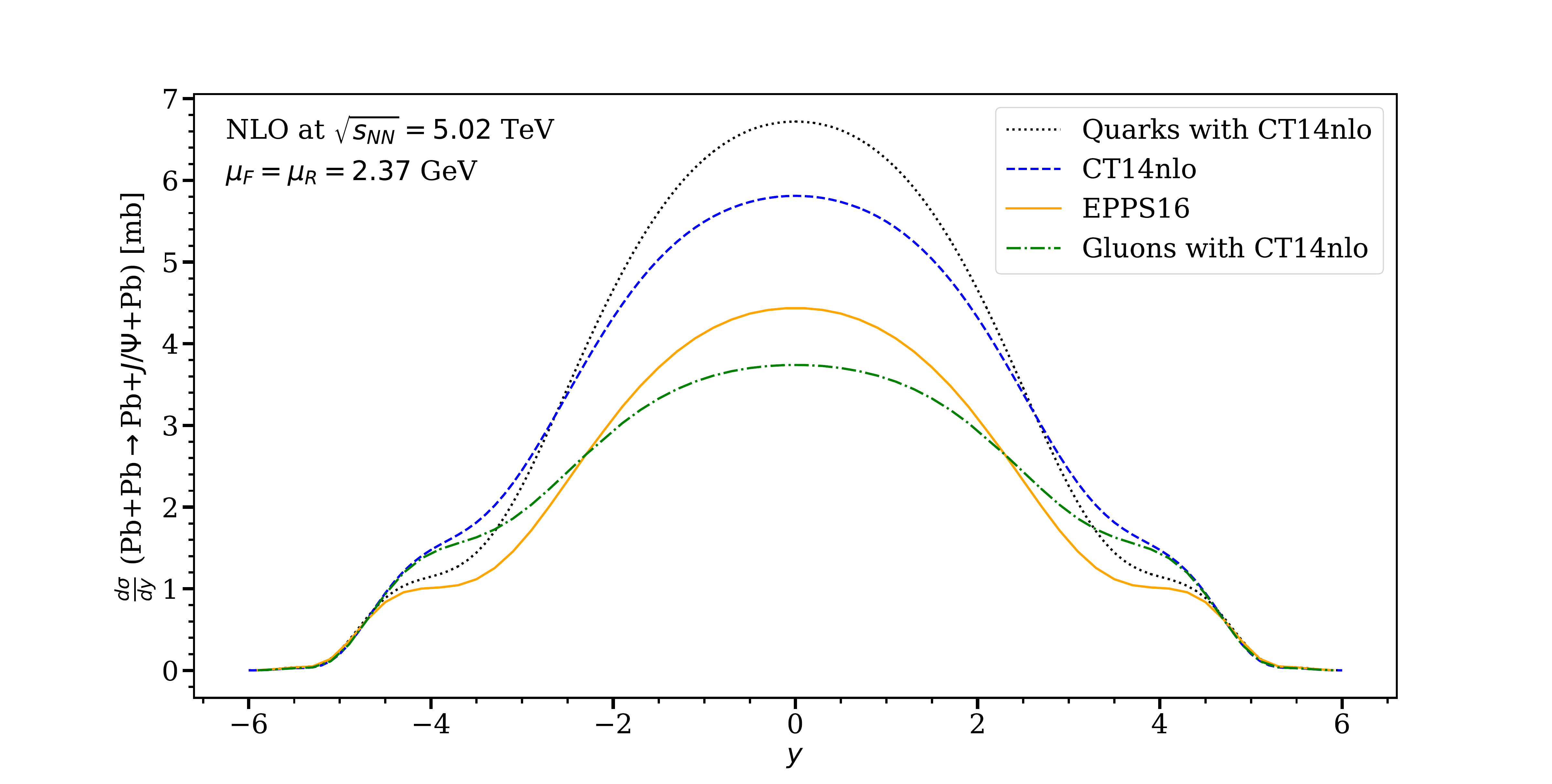}
\caption{Rapidity-differential exclusive $J/\psi$ photoproduction cross section in 5.02 TeV Pb+Pb UPCs, computed with the EPPS16 nPDFs (solid orange curve), with CT14NLO PDFs (dashed blue curve), with CT14NLO gluons and EPPS16 quarks (dotted-dashed green curve), and with CT14NLO quarks and EPPS16 gluons. Notice that turning off the nuclear effects in gluons reduces the cross section at $y=0$ -- for explanation, see the text.
}
    \label{fig:NLO_NuclearSensitivity}
\end{figure*}

Next, we analyse how the nuclear modifications of the PDFs as well as the uncertainties of the nuclear and free-proton PDFs propagate into the exclusive rapidity-differential $J/\psi$ photoproduction cross sections.  
Figure~\ref{fig:NLO_NuclearSensitivity} compares the rapidity-differential cross sections at 5.02 TeV obtained at our ``optimal" scale with the EPPS16 nPDFs (solid orange curve), and the one obtained with the CT14NLO free-proton PDFs (dashed blue) which are the baseline for EPPS16. As seen in the figure, at mid-rapidity, where the $W^\pm$ terms contribute equally, the cross sections show a reduction of a factor of 0.76 from CT14NLO to EPPS16. Towards backward/forward rapidites, i.e. in the regions where the $W^\pm$ terms contribute significantly and probe the nuclear effects in different $x$-regions, the net nuclear effects are slighly increasing. Finally at the backwardmost (forwardmost) rapidities, where the single $W^+$ $(W^-)$ contribution dominates and one enters the antishadowing region, the nuclear effects essentially die out.  

The general behaviour and magnitude of the nuclear effects here can be understood as follows: 

\textbullet~ First, we recall from Figs.~\ref{fig:NLO_EPPS_oGoQ} and \ref{fig:Re_Im_EPPS16}
that it is the imaginary part of the quark amplitude that dominates the cross section at $y=0$. Recalling that $\xi(y) = \zeta (y)/(2-\zeta (y))$, where $\zeta (y) = M_{J/\psi}^2/W^2$ and $W^2 = M_{J/\psi}e^{y}\sqrt{s_{\rm NN}}$, we have $\xi(y=0)\approx 3\cdot 10^{-4}$. This is deep in the shadowing region of nPDFs, and in EPPS16 at this $x$ and our ``optimal" scale the average nuclear sea-quark (gluon) modification is about  0.68 (0.74). The fact that  there seems to be a weaker than quadratic dependence on the PDF's nuclear modification factor follows to our understanding from two reasons: First, in the NLO amplitudes one integrates the parton distributions over $x$ from zero to one: At $x\lesssim\xi$ shadowing is about a constant factor (in EPPS16) while at $x\gtrsim\xi$ shadowing diminishes, so that the net effect of the $x$-integration is a taming of the nuclear effect from that at $x=\xi$. Second, and most importantly, as discussed in detail below, it is again the surprisingly complicated interplay of the different parts of the amplitude and in particular the mutual cancellation of the LO and NLO gluon amplitudes that causes the quark-gluon mixing term to actually cancel some of the nuclear effects. 

\textbullet~ Towards backward rapidities there are competing nuclear effects as $W^+$ decreases, the probed values of $\xi$ increase and the nuclear modifications thereby decrease, and as simultaneously $W^-$ increases, the probed values of  $\xi$  decrease and the nuclear modifications thereby increase (and towards forward rapidities conversely). And, as seen in Fig.~\ref{fig:NLO_EPPS_oGoQ} also quarks and gluons compete over the dominance of cross section, the quark dominance turning into a gluon one towards backward/forward rapidities.

\textbullet~ At the backward rapidity $y=-4$ then, we recall the $W^+$ contribution (Fig.~\ref{fig:NLO_PlusMinDecomp}) and the NLO real part  of the full amplitude (Figs.~\ref{fig:LOandNLOReIm} and \ref{fig:Re_Im_EPPS16}) to dominate the cross section, and from Fig.~\ref{fig:NLO_EPPS_oGoQ} we again see that both quarks and gluons contribute here. Now $\xi(y=-4)\approx 1.7\cdot10^{-2}$ and the EPPS16 gluon (quark) modification is a factor of 0.88 (0.86) while the net nuclear effect is about a factor of 0.68, i.e. surprisingly large. In this region the integration over $x$ does not tame the nuclear effects to the same degree as at small values of $x$, and in particular the large and negative quark-gluon mixing term drives the efficiency of  nuclear effects up here.
 
Given the complex intertwined structure of the cross section, it is also useful to analyse what happens if  we start from the EPPS16 result and turn separately off the nuclear effects from gluons and quarks, one at the time. 

\textbullet~ Turning first off the nuclear effects (suppression) in the gluon PDFs results in the dashed-dotted (green) curve labelled ``Gluons with CT14NLO" in Fig.~\ref{fig:NLO_NuclearSensitivity}, which shows a \textit{reduction} in the cross section relative to the EPPS16 result (solid orange curve) at mid-rapidity. This seems again quite counter-intuitive, as we would naively expect a removal of suppression to cause an \textit{increase} instead.  
Such a behaviour can, however, be again understood by studying the real and imaginary parts of the amplitude: In their absolute values, $\text{Re}(\mathcal{M}_G^\text{LO})$, $\text{Re}(\mathcal{M}_G^\text{NLO})$, $\text{Im}(\mathcal{M}_G^\text{LO})$ and $\text{Re}(\mathcal{M}_G^\text{NLO})$ all behave as expected, i.e. their absolute values indeed \textit{grow} when the nuclear shadowing (suppression) is removed. However, nuclear modifications of the PDFs affect the LO and NLO amplitudes in a slightly different manner. Hence, the degree of the cancellation of $\text{Re}(\mathcal{M}_G^\text{LO})$ against $\text{Re}(\mathcal{M}_G^\text{NLO})$, and $\text{Im}(\mathcal{M}_G^\text{LO})$ against $\text{Im}(\mathcal{M}_G^\text{NLO})$ changes when switching the PDFs from EPPS16 to CT14NLO. With the CT14NLO gluons at this scale, the cancellation of $\text{Im}(\mathcal{M}_G^\text{LO})$ against $\text{Im}(\mathcal{M}_G^\text{NLO})$ happens to be practically perfect. This in turn eliminates the previously large contribution $2[\text{Im}(\mathcal{M}_G^\text{LO})+\text{Im}(\mathcal{M}_G^\text{NLO})]\text{Im}(\mathcal{M}_Q^\text{NLO})$ in the quark-gluon mixing term, causing the \textit{suppression} that we see in Fig.~\ref{fig:NLO_NuclearSensitivity} at mid-rapidity.

\textbullet~ Then, turning off the nuclear effects in the quark distributions, but leaving them on in the gluon contribution results in the dotted black curve which lies rather close to the  
pure CT14NLO case of no nuclear PDF effects at all. This time this is an obvious result, as at mid-rapidity the quark part $\text{Im}(\mathcal{M}_Q^\text{NLO})$ dominates the cross section and removing the suppression in the PDFs just increases the cross section as expected. Figure~\ref{fig:NLO_NuclearSensitivity} thus underlines the quark dominance demonstrated earlier in Fig.~\ref{fig:NLO_EPPS_oGoQ}.

\begin{figure*}[ht]
    \centering
    \includegraphics[width=.8\textwidth]{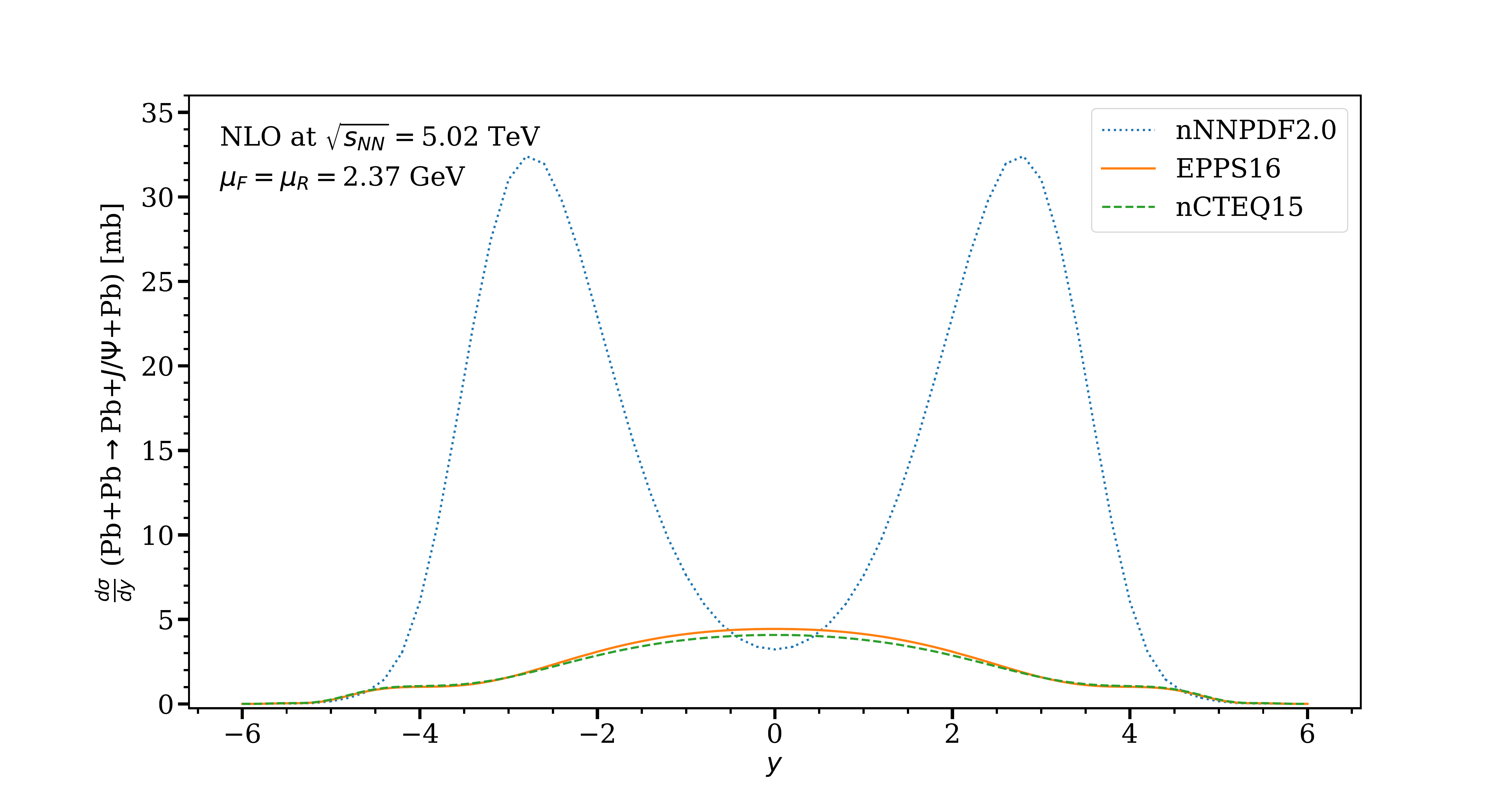}
    \caption{Rapidity-differential exclusive photoproduction of $J/\psi$ in 5.02 TeV Pb+Pb UPCs, computed at our ``optimal" scale using the 
		EPPS16~\cite{Eskola:2016oht} (solid orange curve), 
		nCTEQ15 \cite{Kovarik:2015cma} (dashed green) and 
		nNNPDF2.0 \cite{AbdulKhalek:2020yuc} (dotted blue) nPDFs.}
    \label{fig:NLO_VarietyAll}
\end{figure*}

Because of the rather counter-intuitive results above, and since there is the integration over $x$ from zero to one in the NLO amplitude, we would like to confirm that NLO exclusive photoproduction of $J/\psi$ in Pb+Pb UPCs at the LHC indeed probes the small-$x$ shadowing region ($x\lesssim 0.03...0.04$ in EPPS16), and not the antishadowing region ($0.03...0.04 \lesssim x \lesssim 0.3$ in EPPS16) in the nPDFs. If the process indeed probes the quark and gluon distributions at  $x= {\cal O}(\xi)$, and  $\xi(y=0) \approx 3\cdot 10^{-4}$, then the biggest effect to the final result (relative to the CT14NLO result above) should be attained by turning on only the nuclear corrections in the shadowing region. We have checked that this is indeed the case: Running the code with ad hoc modified nPDFs that coincide with EPPS16 in the shadowing region and with CT14NLO elsewhere, the results are essentially (within 6\%) the same as the EPPS16 results. 

Next, we investigate how sensitive the studied cross sections are to the choice of the nPDFs. Figure~\ref{fig:NLO_VarietyAll} shows the rapidity-differential cross sections obtained with the central sets of the EPPS16 (solid orange curve), nCTEQ15  (dashed green) and  nNNPDF2.0 (dotted blue) nPDFs.
The nCTEQ15 set gives essentially the same result as EPPS16 but there seems to be a huge difference to the nNNPDF2.0 set. The shape of the nNNPDF2.0 result is very different from EPPS16/nCTEQ15, and the magnitude at forward and backward rapidities is off by about a factor of 15. We have traced the very fast growth of the cross section down to the rapidly growing real part of the LO gluon amplitude, which includes again the integration over $x$ from 0 to 1 where the small-$x$ gluons (in the ERBL region $x\lesssim\xi$ but near $x\sim \xi$) start to play a significant role with nNNPDF2.0. The real part of the LO gluon amplitude is not as well numerically cancelling against the real part of the NLO gluon amplitude with nNNPDF2.0 as with EPPS16/nCTEQ15, which  in turn makes the forward/backward-$y$ cross section again more sensitive to the small-$x$ gluon distributions, and this is what we see in Fig.~\ref{fig:NLO_VarietyAll}. 

\begin{figure*}[ht]
    \centering
    \includegraphics[width=.8\textwidth]{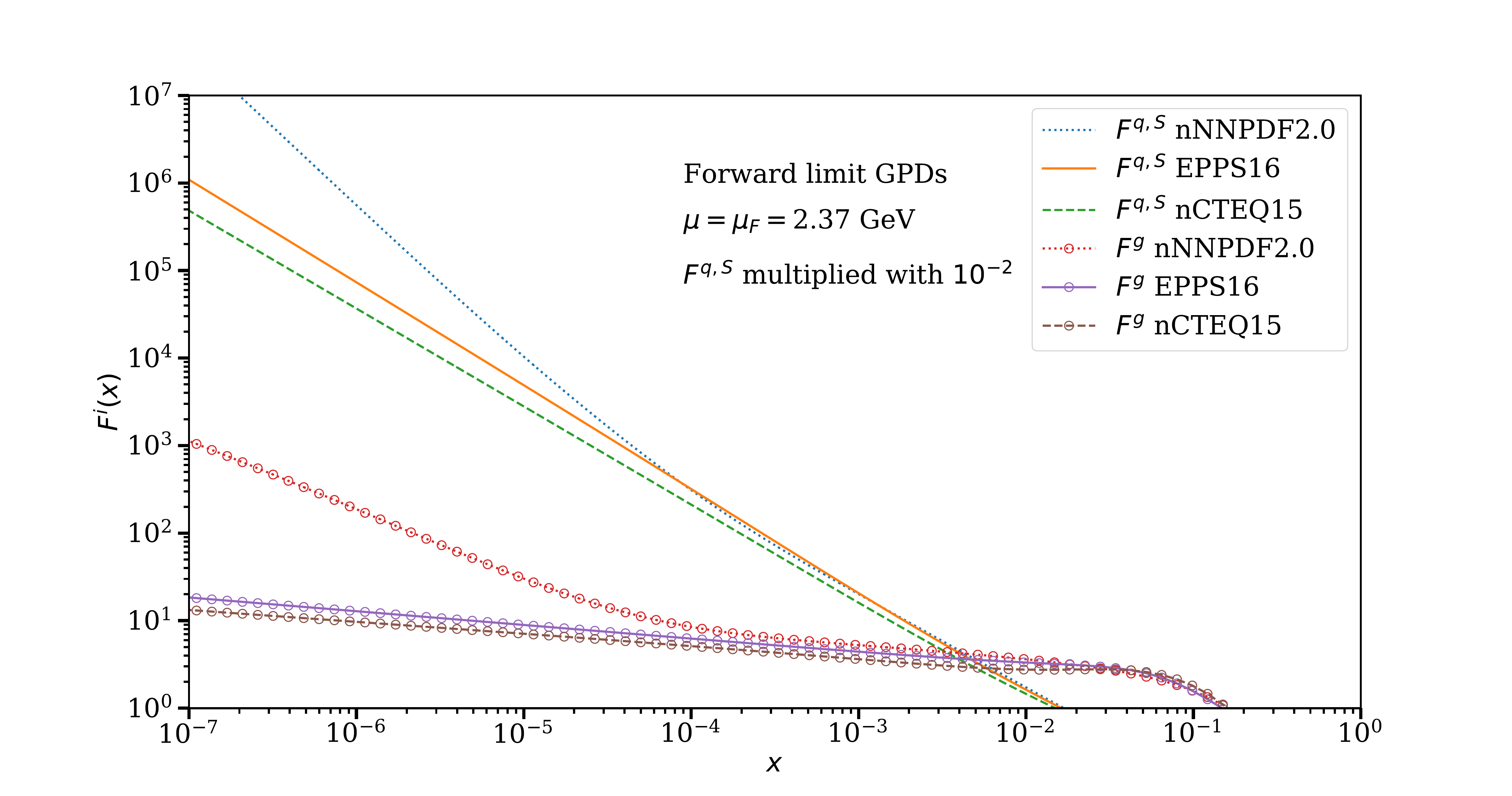}
    \caption{The nPDF gluon distributions $xg(x,\mu)$ and the quark singlet distributions $F^{q,S}=\sum_q[q(x,\mu)+\bar q(x,\mu)]$ as given by EPPS16 (solid lines), nCTEQ15 (dashed) and nNNPDF2.0 (dotted) nPDFs at the ``optimal" scale. }
    \label{fig:nPDF_Variety}
\end{figure*}

We plot in Fig.~\ref{fig:nPDF_Variety} the gluon distributions $xg(x,\mu)$ and the quark singlet distributions $F^{q,S}=\sum_q[q(x,\mu)+\bar q(x,\mu)]$ from EPPS16, nCTEQ15 and nNNPDF2.0 nPDFs as they enter our computation at the ``optimal" scale. The figure confirms the similarity of the EPPS16 and nCTEQ15 PDFs and shows that the nNNPDF2.0 quarks differ systematically from these at $x\lesssim 10^{-5}$ and the gluons at $x\lesssim10^{-4}$.  In Fig.~\ref{fig:NLO_VarietyAll}, the increased small-$x$ gluons of nNNPDF2.0 make the $W^-$ component of the cross section to be the dominant one at $y=-3$. For the $W^-$ contribution $\xi(y=-3)={\cal O}(10^{-5})$, and at these values of $x$, Fig.~\ref{fig:nPDF_Variety} indicates already a factor of three difference between the nNNPDF2.0 and EPPS16/nCTEQ15 gluons. The square of this difference then explains the order of magnitude of the difference between the nNNPDF and EPPS16/nCTEQ15 results seen in  Fig.~\ref{fig:NLO_VarietyAll}. 

\begin{figure*}[ht]
    \centering
    \includegraphics[width=.8\textwidth]{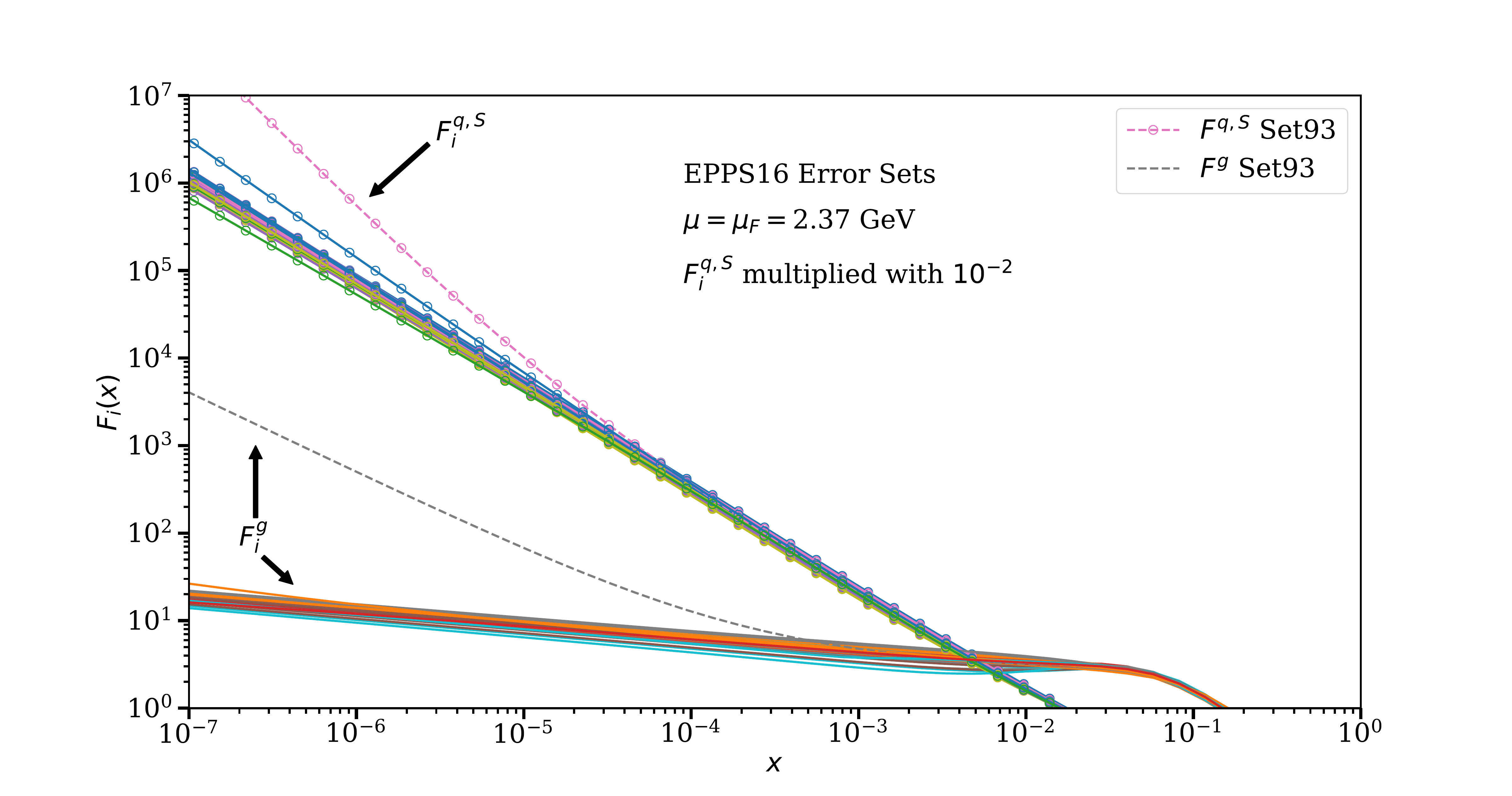}
    \caption{The error sets of EPPS16\&CT14NLO nPDFs\&PDFs for the gluon (lower set of curves) and quark-singlet distributions (upper set) as functions of $x$, at our ``optimal" scale. Altogether 96 error sets are plotted, of which the numbers 1-40 in the LHAPDF set-up \cite{Buckley:2014ana} of EPPS16 are for the nuclear effects and 	41-96 for the CT14NLO free-proton PDFs. The CT14NLO-related Set93 is the one clearly standing out from the rest at $x\lesssim10^{-4}$.}
    \label{fig:EPPS16_Errors}
\end{figure*}

Next, we investigate the PDF uncertainties in the computed rapidity-differential cross sections and compare them with the existing data. We propagate the PDF/nPDF uncertainties to the computed cross sections using the asymmetric form \cite{Eskola:2016oht}
\begin{equation}
    \delta \mathcal{O}^{\pm} = \sqrt{\sum\limits_i [^{\text{max}}_{{\text{min}}} \left\{ \mathcal{O}(S_i^+)- \mathcal{O}(S_0), \mathcal{O}(S_i^-)- \mathcal{O}(S_0),0 \right\} ]^2 },
\end{equation}
where $S_i^{\pm}$ labels the error sets for the given PDF. We plot the error sets of EPPS16+CT14NLO in Fig.~\ref{fig:EPPS16_Errors} for the gluon distributions $xg(x,\mu)$, and for the quark singlet distributions $F^{q,S}$, again at our ``optimal scale". As the figure shows, one CT14-related error set, ``Set93", of the EPPS16 implementation in LHAPDF \cite{Buckley:2014ana} (error set 53 in CT14NLO), stands clearly out at smallest values of $x$, and even more strongly than the nNNPF2.0 PDFs did in Fig.~\ref{fig:nPDF_Variety}, while the rest of the EPPS16-related and CT14-related error sets show only rather moderate variations w.r.t. the central sets. Similarly to the case with the nNNPDF2.0 nPDFs above, the rapid growth of the small-$x$ gluon distributions in this error set induces again a rapid growth of the real part of the LO gluon amplitude, and hence the cross sections.

\begin{figure*}[t]
    \centering
    \includegraphics[width=.8\textwidth]{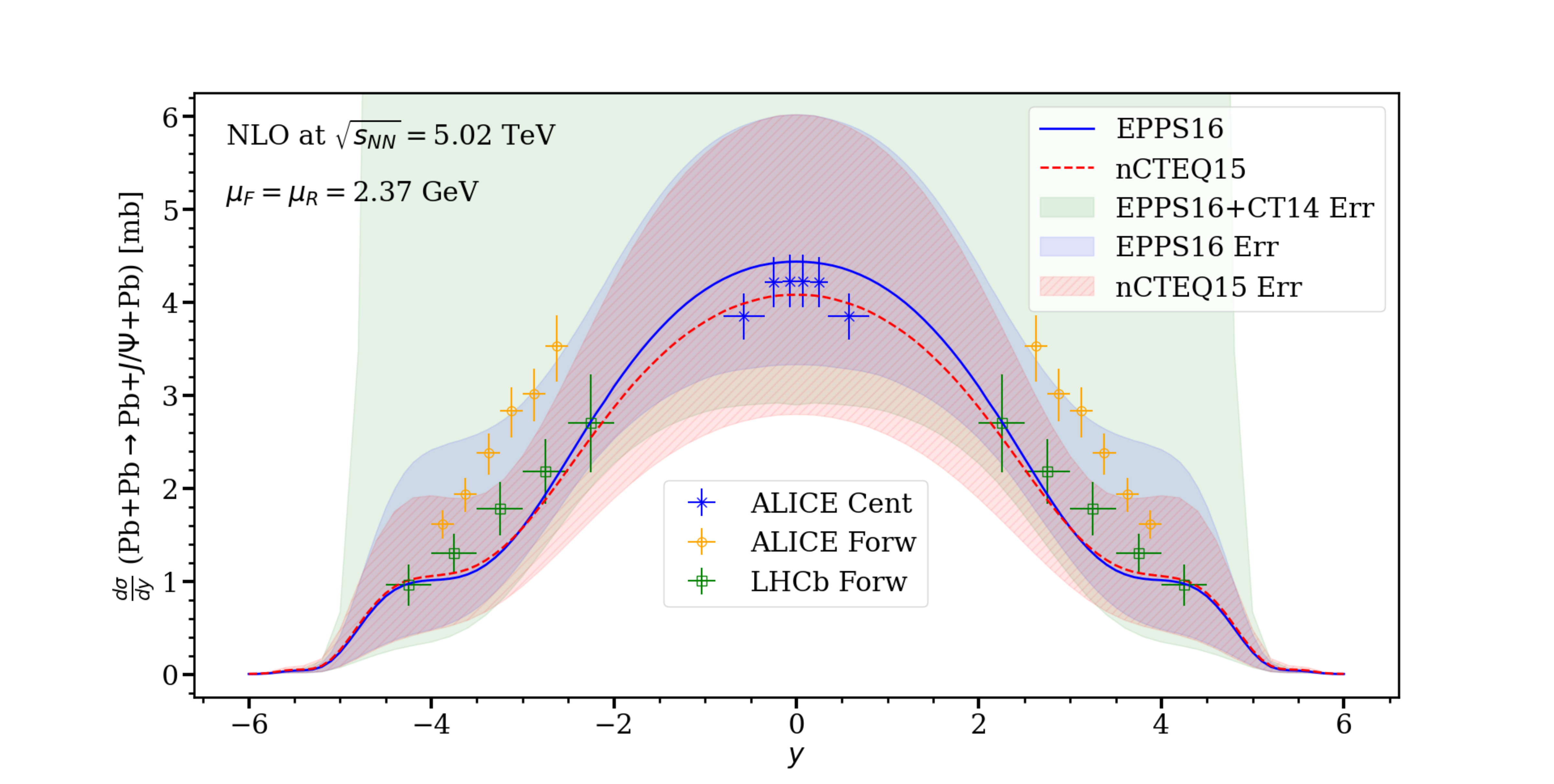}
    \includegraphics[width=.8\textwidth]{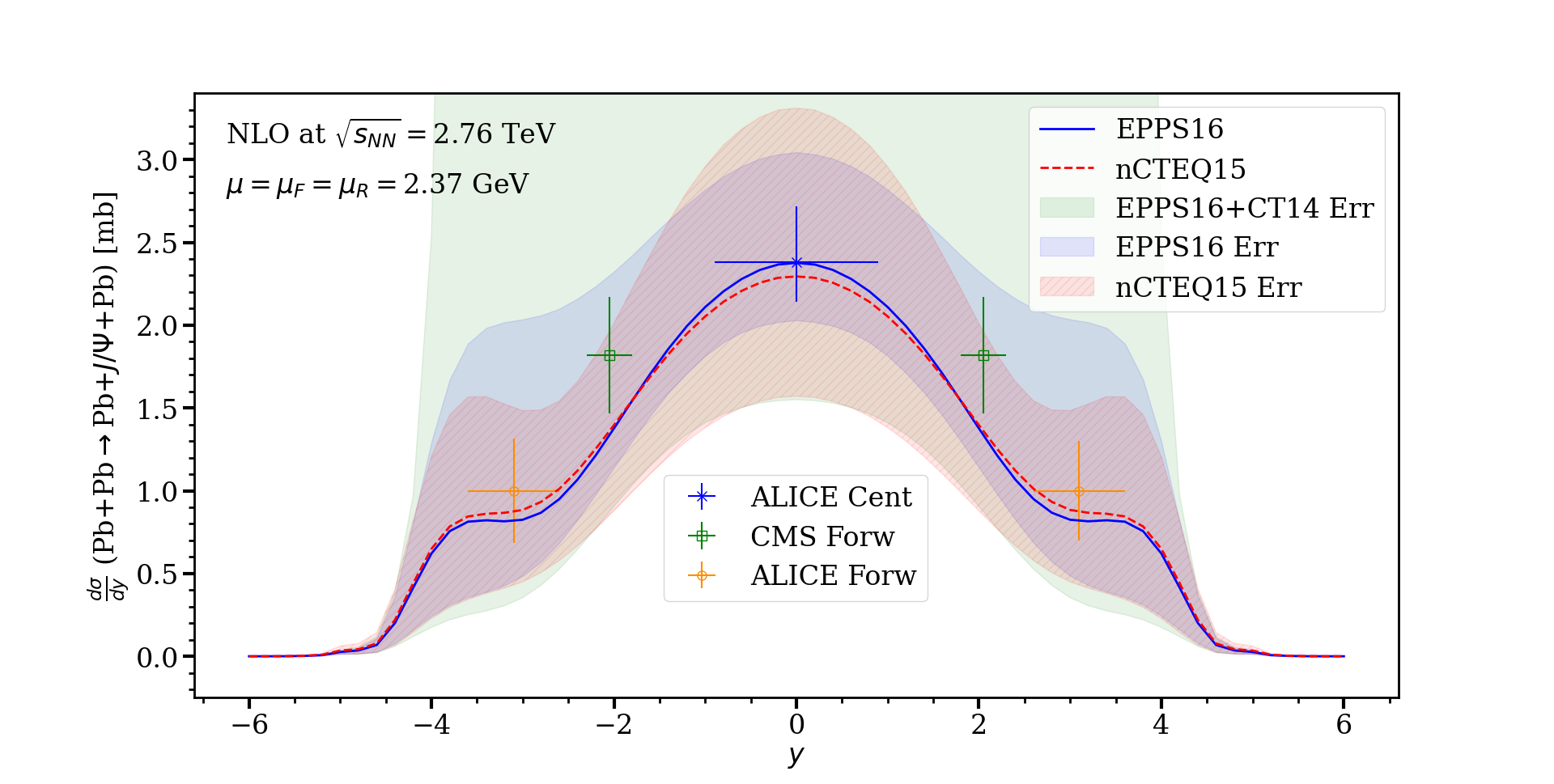}
\caption{\textbf{Upper panel:} Uncertainties originating from the nPDFs/PDFs in the rapidity-differential exclusive $J/\psi$ photoproduction NLO cross sections in 5.02 TeV Pb+Pb UPCs, computed at our ``optimal" scale $\mu = 2.37$~GeV using the EPPS16+CT14NLO and nCTEQ15 error sets. The solid (dashed) line shows the EPPS16+CT14NLO (nCTEQ15) central-set result, and the corresponding uncertainty bands are explained in the text. The experimental datapoints are from Run2 and the same as in the upper panel of Fig.~\ref{fig:Run1Run2Envelope}. 
\textbf{Lower panel:} The same but for $\sqrt{s_{\rm NN}}=2.76$~TeV and with the same Run1 data as in the lower panel of Fig.~\ref{fig:Run1Run2Envelope}.} \label{fig:276-502-AllErrors}
\end{figure*}    

Figure~\ref{fig:276-502-AllErrors} shows the uncertainties that are induced to the rapidity-differential exclusive $J/\psi$ photoproduction cross sections in 5.02~TeV (upper panel) and 2.76~TeV (lower panel) Pb+Pb UPCs by the PDF/nPDF uncertainties. The uncertainties arising from the EPPS16 nuclear effects alone are shown by the dark (blue) bands, while the full EPPS16+CT14NLO error bands (green) contain uncertainties from both the nuclear effects and the free-proton baseline PDFs. The results with EPPS16 and CT14NLO central sets are shown by the solid (blue) curves. As expected based on Fig.~\ref{fig:EPPS16_Errors}, ``Set93" above entirely dictates the green error bands. The EPPS16+CT14NLO full uncertainty band at mid-rapidity (not shown in the figure) goes up to some   150 (37) mb and at $y \approx \pm 2.2$  as high as 1500 (170) mb for the  5.02 (2.76)~TeV collision energy. We also have checked that without ``Set93" the CT14NLO uncertainties become of the same order and slightly smaller than those for EPPS16. For comparison with the EPPS16 results, we also plot the uncertainty bands (hatched) arising from the nCTEQ15 error sets. These now account for the uncertainties in the nuclear effects only, and not in the free-proton PDFs. The central-set results with nCTEQ15 are shown by the dashed (red) line. We should also emphasize that the nCTEQ15 results here have been obtained at our ``optimal" scale, without further tuning of the scale.  

As we have already seen, the EPPS16 results produce a relatively good fit to the experimental Run1 and Run2 data at our ``optimal" scale, and as seen in Fig.~\ref{fig:276-502-AllErrors}, so do the nCTEQ15 ones, too.  The uncertainties arising from the nuclear effects in EPPS16 and nCTEQ15 are of the same order of magnitude mutually, and typically somewhat larger than the errorbars of the data. As the figure indicates, one must not forget the free-proton PDF uncertainties when considering absolute cross sections. Finally, regarding the tension between the ALICE and LHCb data in the forward/backward direction, we can see that at least at our ``optimal" scale both the EPPS16 and nCTEQ15 results (but obviously not the nNNPDF2.0) seem to reproduce the LHCb data points better but that both data sets can still be accommodated within the larger EPPS16 uncertainties.

\section{Summary} \label{Sec:Summary}

We have presented the very first implementation of exclusive rapidity-differential $J/\psi$ photoproduction cross sections in ultraperipheral nucleus-nucleus collisions in the framework of  collinear factorization and NLO perturbative QCD. We have developed our numerical code for the ultraperipheral nuclear collisions  based on the analytical NLO results of Ref.~\cite{Ivanov:2004vd}, utilizing the experience obtained also in \cite{Jones:2013pga,Flett:2021xsl}, and following earlier literature in accounting for the photon fluxes of the colliding nuclei 
\cite{vonWeizsacker:1934nji, Vidovic:1992ik, Jackson:1998nia,Baltz:2007kq,Adeluyi:2012ph,Guzey:2013taa,Klein:2016yzr,Zha:2018ywo} and for the $t$-dependence of the cross section with a standard nuclear form factor. In this exploratory NLO study for the UPCs, we approximate the GPDs involved in the process with their forward-limit nuclear PDFs. Our default choice for the nPDFs and their error sets is EPPS16 \cite{Eskola:2016oht} but we also study the nPDF sensitivity of our results by using nCTEQ15 \cite{Kovarik:2015cma} and nNNPDF2.0 \cite{AbdulKhalek:2020yuc}. 

We have shown that, as expected based on Ref.~\cite{Ivanov:2004vd}, the computed rapidity-differential NLO cross sections of $J/\psi$ photoproduction in 5.02 and 2.76 TeV Pb+Pb UPCs at the LHC, as well as the corresponding photoproduction cross sections in $ep$ collisions at HERA, are both in their magnitude and in their shape quite sensitive to the scale choice. As the scale-sensitivity is much larger than the error bars of the experimental data at the LHC, it makes it difficult to make solid NLO predictions of the corresponding $J/\psi$ cross section for UPCs at other energies. Quite encouragingly, however, we have found that a scale-choice $\mu \approx 0.76 M_{J/\psi}$, which lies in the physically reasonable range $\mu = {\cal O}(M_{J/\psi})$, can actually be determined, with which we can well reproduce the ALICE \cite{ALICE:2019tqa,ALICE:2012yye,ALICE:2021gpt,ALICE:2013wjo}, LHCb \cite{LHCb:2021bfl} and CMS \cite{CMS:2016itn} UPC data at these energies. We have also tested that the same scale choice, called here ``optimal" scale, works well also with the nCTEQ15 nPDFs. Interestingly, in studying the scale-sensitivity at a fixed value of $y=0$, we noticed that towards the upper end of the scales studied here the scale sensitivity of the full NLO result becomes actually weaker than that of the LO result, but towards the lower end of scales stronger than in LO. Also interestingly, at midrapidity the ``optimal" scale becomes fixed right in the scale-region where the NLO contributions are the smallest relative to LO. In the future, it will be interesting to see whether this ``minimal-sensitivity" feature remains there also after further modeling of the GPDs. 

We have made an effort to analyse in sufficient detail the surprisingly complex structure of the exclusive rapidity-differential $J/\psi$ photoproduction NLO cross sections in Pb+Pb UPCs at 5.02 and 2.76 TeV. In particular, we have shown how the computed NLO cross sections form under various competing and intertwining effects: There are competing contributions from the photon-nucleon c.m.s. energy $W^\pm$ components,  from the real and imaginary parts of the full amplitude, from the quark and gluon GPD/PDF contributions which also mix in a non-trivial way in the squared amplitude, and most importantly of all, from the gluonic LO and NLO amplitudes which come with opposite signs and cancel each other to a degree that nontrivially depends on the $W^\pm$. All these competing contributions need to be taken into account in the full NLO study, as is done in the current paper. 

The main result of our NLO study with the EPPS16 nPDFs, similar to the findings in Ref.~\cite{Ivanov:2004vd} but now for UPCs, is that due to the cancelling LO and NLO gluon amplitudes it is predominantly the small-$x$ \textit{quark} GPDs/PDFs that exclusive $J/\psi$ photoproduction is probing in UPCs at midrapidity, and not the gluon distributions as has been traditionally suggested before based on LO. This is an important result not addressed to our knowledge in the UPC context before. We have also checked that this result is robust against the scale variation studied here. We have also shown that towards the forward/backward rapidities the gluon dominance is eventually recovered but because of the folding with the photon flux (which kills one of the $W^\pm$ contributions) the nuclear gluon GPDs/PDFs become probed at larger values of $x$ (where shadowing effects become smaller) than at midrapidity. Thus, our conclusion is  -- at least in our current ``bare bones" GPD/PDF framework and with the EPPS16 and nCTEQ15 nPDFs -- that the exclusive rapidity-differential $J/\psi$ photoproduction cross sections at the LHC are not as a direct and efficient probe of the small-$x$ nuclear gluon PDFs as thought before, but that they are primarily probed (at midrapidity at least) through the DGLAP evolution of the quark GPDs/PDFs. Another important observation is that at midrapidity the dependence of the computed NLO cross sections on the nuclear effects in PDFs is not as quadratic as thought in the LO gluon context before. The taming of the net nuclear effects follows partly from the $x$-integration in the NLO amplitude but predominantly from the behaviour of the interference term in the squared amplitude which mixes the quark and gluon contributions in a non-trivial way.

We have also investigated the dependence of our results on the uncertainties of the PDFs. The nCTEQ15 central-set results are essentially the same as those with the central set of EPPS16. At midrapidity, where the quark contributions dominate, these two sets show very similar error bands when the uncertainties of the nuclear effects in the PDFs are propagated into the NLO cross sections. Towards forward/backward rapidities where gluons dominate, the EPPS16 uncertainties become slightly larger, which follows from the more realistic (due to having more freedom in the gluon PDF shape there) estimates of the gluon nPDF uncertainties than in nCTEQ15. In any case, in the current ``bare bones" GPD/PDF framework, we observe that both the forward ALICE \cite{ALICE:2019tqa} and LHCb \cite{LHCb:2021bfl} data can be accommodated within the nuclear PDF error bands, while the results with the central sets of EPPS16 and nCTEQ15 agree better with the LHCb data.  

Finally, we have observed that if there is a very rapid rise in the small-$x$  gluon distributions, such as in the nNNPDF2.0 central set and the error set 53 in the CT14NLO free-proton PDFs \cite{Dulat:2015mca} (93 in EPPS16 at LHAPDF), then the smallest-$x$ contribution to the real part of the gluon LO  amplitude starts to dominate the cross sections. Concretely, in our results when the EPPS16 nuclear errors and the CT14NLO errors are appropriately combined, the CT14NLO error set 53 (93 in EPPS16 at LHAPDF) dictates the upper boundaries of the very large uncertainty band on our central result. In our ``bare bones" GPD/PDF framework, such a growth seems to be ruled out by the UPC data considered here. However, before we can make any further conclusions on this point, uncertainties arising from the modeling of GPDs should be quantified. 

The current paper is meant as a baseline for systematic further studies of exclusive photoproduction of vector mesons in ultraperipheral nucleus-nucleus collisions, in collinear factorization and NLO pQCD. An obvious next task is to repeat the NLO study for the photoproduction cross sections of $\Upsilon$ mesons, to investigate in particular how much the scale dependence changes and check exactly what happens with all the intertwined effects at the higher scales $\mu={\cal O}(M_\Upsilon)$. On the basis of Ref.~\cite{Ivanov:2004vd}, we would expect to see a reduced scale-sensitivity and a stronger dependence on the gluon PDFs also in the UPC case.

There are also several ways the current framework could and should be improved.
Our strategy for the current exploratory study is that as the scale- and PDF-related uncertainties are so large, we may leave the GPD modeling (such as in Ref.~\cite{Freund:2002qf}) as a future challenge. Next, given the studied ``bare bones" GPD/PDF baseline, it will be interesting to study how the nPDF uncertainties propagate to the GPDs and via them to the NLO cross sections. As far as we can see, based e.g. on Refs. \cite{Martin:1997wy,Shuvaev:1999ce}  the skewedness corrections to the GPD quark distributions in the DGLAP region can be expected to be larger for quarks than for gluons, which would further strengthen our conclusion of the quark dominance at mid-rapidity. Towards forward/backward rapidities, the gluon dominance would then correspondingly kick in more slowly. Particularly interesting here would be to study the role of the nuclear effects in the ERBL region, where the PDFs are known not to be an optimal approximation but which in the current study turned out to be important essentially only with PDF sets that have rapidly growing small-$x$ distributions. Future improvements would also include non-relativistic QCD corrections into the vector meson wavefunction \cite{Hoodbhoy:1996zg,Frankfurt:1997fj,Lappi:2020ufv}.

\vspace{0.5cm}

\textit{Acknowledgements} 
We acknowledge the helpful discussions with I.~Helenius and H.~M\"antysaari. We acknowledge the financial support from the Magnus Ehrnrooth foundation (T.L.), the Academy of Finland projects 297058 (K.J.E.), 308301 (H.P.) and 330448 (K.J.E.). This research was funded as a part of the Center of Excellence in Quark Matter of the Academy of Finland (project 346325). This research is part of the European Research Council project ERC-2018-ADG-835105 YoctoLHC.

\bibliographystyle{JHEP-2modlong.bst}
\bibliography{apssamp}

\providecommand{\noopsort}[1]{}\providecommand{\singleletter}[1]{#1}%
\providecommand{\href}[2]{#2}\begingroup\raggedright\begin{thebibliography}{10}

\bibitem{Bertulani:1987tz}
C.~A. Bertulani and G.~Baur, {\it {Electromagnetic Processes in Relativistic
  Heavy Ion Collisions}},
  \href{http://dx.doi.org/10.1016/0370-1573(88)90142-1}{{\em Phys. Rept.} {\bf
  163} (1988) 299}.

\bibitem{Nystrand:2006gi}
J.~Nystrand, {\it {Ultra-peripheral collisions of heavy ions at RHIC and the
  LHC}},  \href{http://dx.doi.org/10.1016/j.nuclphysa.2006.12.015}{{\em Nucl.
  Phys. A} {\bf 787} (2007) 29}
  [\href{http://arXiv.org/abs/hep-ph/0611042}{{\tt arXiv:hep-ph/0611042}}].

\bibitem{Baltz:2007kq}
A.~J. Baltz, {\it {The Physics of Ultraperipheral Collisions at the LHC}},
  \href{http://dx.doi.org/10.1016/j.physrep.2007.12.001}{{\em Phys. Rept.} {\bf
  458} (2008) 1} [\href{http://arXiv.org/abs/0706.3356}{{\tt arXiv:0706.3356
  [nucl-ex]}}].

\bibitem{Adeluyi:2011rt}
A.~Adeluyi and C.~Bertulani, {\it {Gluon distributions in nuclei probed at the
  CERN Large Hadron Collider}},
  \href{http://dx.doi.org/10.1103/PhysRevC.84.024916}{{\em Phys. Rev. C} {\bf
  84} (2011) 024916} [\href{http://arXiv.org/abs/1104.4287}{{\tt
  arXiv:1104.4287 [nucl-th]}}].

\bibitem{Adeluyi:2012ph}
A.~Adeluyi and C.~A. Bertulani, {\it {Constraining Gluon Shadowing Using
  Photoproduction in Ultraperipheral pA and AA Collisions}},
  \href{http://dx.doi.org/10.1103/PhysRevC.85.044904}{{\em Phys. Rev. C} {\bf
  85} (2012) 044904} [\href{http://arXiv.org/abs/1201.0146}{{\tt
  arXiv:1201.0146 [nucl-th]}}].

\bibitem{Guzey:2016piu}
V.~Guzey, E.~Kryshen and M.~Zhalov, {\it {Coherent photoproduction of vector
  mesons in ultraperipheral heavy ion collisions: Update for run 2 at the CERN
  Large Hadron Collider}},
  \href{http://dx.doi.org/10.1103/PhysRevC.93.055206}{{\em Phys. Rev. C} {\bf
  93} (2016)~no.~5 055206} [\href{http://arXiv.org/abs/1602.01456}{{\tt
  arXiv:1602.01456 [nucl-th]}}].

\bibitem{Guzey:2020ntc}
V.~Guzey, E.~Kryshen, M.~Strikman and M.~Zhalov, {\it {Nuclear suppression from
  coherent $J /\psi$ photoproduction at the Large Hadron Collider}},
  \href{http://dx.doi.org/10.1016/j.physletb.2021.136202}{{\em Phys. Lett. B}
  {\bf 816} (2021) 136202} [\href{http://arXiv.org/abs/2008.10891}{{\tt
  arXiv:2008.10891 [hep-ph]}}].

\bibitem{Guzey:2013xba}
V.~Guzey, E.~Kryshen, M.~Strikman and M.~Zhalov, {\it {Evidence for nuclear
  gluon shadowing from the ALICE measurements of PbPb ultraperipheral exclusive
  $J/{\psi}$ production}},
  \href{http://dx.doi.org/10.1016/j.physletb.2013.08.043}{{\em Phys. Lett. B}
  {\bf 726} (2013) 290} [\href{http://arXiv.org/abs/1305.1724}{{\tt
  arXiv:1305.1724 [hep-ph]}}].

\bibitem{Guzey:2013qza}
V.~Guzey and M.~Zhalov, {\it {Exclusive $J/{\psi}$ production in
  ultraperipheral collisions at the LHC: constrains on the gluon distributions
  in the proton and nuclei}},
  \href{http://dx.doi.org/10.1007/JHEP10(2013)207}{{\em JHEP} {\bf 10} (2013)
  207} [\href{http://arXiv.org/abs/1307.4526}{{\tt arXiv:1307.4526 [hep-ph]}}].

\bibitem{Guzey:2016qwo}
V.~Guzey, M.~Strikman and M.~Zhalov, {\it {Accessing transverse nucleon and
  gluon distributions in heavy nuclei using coherent vector meson
  photoproduction at high energies in ion ultraperipheral collisions}},
  \href{http://dx.doi.org/10.1103/PhysRevC.95.025204}{{\em Phys. Rev. C} {\bf
  95} (2017)~no.~2 025204} [\href{http://arXiv.org/abs/1611.05471}{{\tt
  arXiv:1611.05471 [hep-ph]}}].

\bibitem{Jones:2016ldq}
S.~P. Jones, A.~D. Martin, M.~G. Ryskin and T.~Teubner, {\it {The exclusive
  $J/\psi$ process at the LHC tamed to probe the low $x$ gluon}},
  \href{http://dx.doi.org/10.1140/epjc/s10052-016-4493-y}{{\em Eur. Phys. J. C}
  {\bf 76} (2016)~no.~11 633} [\href{http://arXiv.org/abs/1610.02272}{{\tt
  arXiv:1610.02272 [hep-ph]}}].

\bibitem{Ryskin:1992ui}
M.~Ryskin, {\it {Diffractive $J/\psi$ electroproduction in LLA QCD}},
  \href{http://dx.doi.org/10.1007/BF01555742}{{\em Z. Phys. C} {\bf 57} (1993)
  89}.

\bibitem{Klein:2016yzr}
S.~R. Klein, J.~Nystrand, J.~Seger, Y.~Gorbunov and J.~Butterworth, {\it
  {STARlight: A Monte Carlo simulation program for ultra-peripheral collisions
  of relativistic ions}},
  \href{http://dx.doi.org/10.1016/j.cpc.2016.10.016}{{\em Comput. Phys.
  Commun.} {\bf 212} (2017) 258} [\href{http://arXiv.org/abs/1607.03838}{{\tt
  arXiv:1607.03838 [hep-ph]}}].

\bibitem{Harland-Lang:2020veo}
L.~A. Harland-Lang, M.~Tasevsky, V.~A. Khoze and M.~G. Ryskin, {\it {A new
  approach to modelling elastic and inelastic photon-initiated production at
  the LHC: SuperChic 4}},
  \href{http://dx.doi.org/10.1140/epjc/s10052-020-08455-0}{{\em Eur. Phys. J.
  C} {\bf 80} (2020)~no.~10 925} [\href{http://arXiv.org/abs/2007.12704}{{\tt
  arXiv:2007.12704 [hep-ph]}}].

\bibitem{Goncalves:2005yr}
V.~P. Goncalves and M.~V.~T. Machado, {\it {The QCD pomeron in ultraperipheral
  heavy ion collisions. IV. Photonuclear production of vector mesons}},
  \href{http://dx.doi.org/10.1140/epjc/s2005-02175-3}{{\em Eur. Phys. J. C}
  {\bf 40} (2005) 519} [\href{http://arXiv.org/abs/hep-ph/0501099}{{\tt
  arXiv:hep-ph/0501099}}].

\bibitem{Lappi:2013am}
T.~Lappi and H.~Mantysaari, {\it {$J/ \psi$ production in ultraperipheral Pb+Pb
  and $p$+Pb collisions at energies available at the CERN Large Hadron
  Collider}},  \href{http://dx.doi.org/10.1103/PhysRevC.87.032201}{{\em Phys.
  Rev. C} {\bf 87} (2013)~no.~3 032201}
  [\href{http://arXiv.org/abs/1301.4095}{{\tt arXiv:1301.4095 [hep-ph]}}].

\bibitem{Mantysaari:2017dwh}
H.~M\"antysaari and B.~Schenke, {\it {Probing subnucleon scale fluctuations in
  ultraperipheral heavy ion collisions}},
  \href{http://dx.doi.org/10.1016/j.physletb.2017.07.063}{{\em Phys. Lett. B}
  {\bf 772} (2017) 832} [\href{http://arXiv.org/abs/1703.09256}{{\tt
  arXiv:1703.09256 [hep-ph]}}].

\bibitem{Mantysaari:2017slo}
H.~M\"antysaari and R.~Venugopalan, {\it {Systematics of strong nuclear
  amplification of gluon saturation from exclusive vector meson production in
  high energy electron\textendash{}nucleus collisions}},
  \href{http://dx.doi.org/10.1016/j.physletb.2018.04.044}{{\em Phys. Lett. B}
  {\bf 781} (2018) 664} [\href{http://arXiv.org/abs/1712.02508}{{\tt
  arXiv:1712.02508 [nucl-th]}}].

\bibitem{Cepila:2017nef}
J.~Cepila, J.~G. Contreras and M.~Krelina, {\it {Coherent and incoherent
  $\mathrm{J/}\psi$ photonuclear production in an energy-dependent hot-spot
  model}},  \href{http://dx.doi.org/10.1103/PhysRevC.97.024901}{{\em Phys. Rev.
  C} {\bf 97} (2018)~no.~2 024901} [\href{http://arXiv.org/abs/1711.01855}{{\tt
  arXiv:1711.01855 [hep-ph]}}].

\bibitem{Mantysaari:2019jhh}
H.~M\"antysaari and B.~Schenke, {\it {Accessing the gluonic structure of light
  nuclei at a future electron-ion collider}},
  \href{http://dx.doi.org/10.1103/PhysRevC.101.015203}{{\em Phys. Rev. C} {\bf
  101} (2020)~no.~1 015203} [\href{http://arXiv.org/abs/1910.03297}{{\tt
  arXiv:1910.03297 [hep-ph]}}].

\bibitem{Sambasivam:2019gdd}
B.~Sambasivam, T.~Toll and T.~Ullrich, {\it {Investigating saturation effects
  in ultraperipheral collisions at the LHC with the color dipole model}},
  \href{http://dx.doi.org/10.1016/j.physletb.2020.135277}{{\em Phys. Lett. B}
  {\bf 803} (2020) 135277} [\href{http://arXiv.org/abs/1910.02899}{{\tt
  arXiv:1910.02899 [hep-ph]}}].

\bibitem{Klein:2019qfb}
S.~R. Klein and H.~M\"antysaari, {\it {Imaging the nucleus with high-energy
  photons}},  \href{http://dx.doi.org/10.1038/s42254-019-0107-6}{{\em Nature
  Rev. Phys.} {\bf 1} (2019)~no.~11 662}
  [\href{http://arXiv.org/abs/1910.10858}{{\tt arXiv:1910.10858 [hep-ex]}}].

\bibitem{Caldwell:2010zza}
A.~Caldwell and H.~Kowalski, {\it {Investigating the gluonic structure of
  nuclei via $J/ \psi$ scattering}},
  \href{http://dx.doi.org/10.1103/PhysRevC.81.025203}{{\em Phys. Rev. C} {\bf
  81} (2010) 025203}.

\bibitem{Bendova:2020hbb}
D.~Bendova, J.~Cepila, J.~G. Contreras and M.~Matas, {\it {Photonuclear
  $J/\psi$ production at the LHC: Proton-based versus nuclear dipole scattering
  amplitudes}},  \href{http://dx.doi.org/10.1016/j.physletb.2021.136306}{{\em
  Phys. Lett. B} {\bf 817} (2021) 136306}
  [\href{http://arXiv.org/abs/2006.12980}{{\tt arXiv:2006.12980 [hep-ph]}}].

\bibitem{Mantysaari:2021ryb}
H.~M\"antysaari and J.~Penttala, {\it {Exclusive heavy vector meson production
  at next-to-leading order in the dipole picture}},
  \href{http://dx.doi.org/10.1016/j.physletb.2021.136723}{{\em Phys. Lett. B}
  {\bf 823} (2021) 136723} [\href{http://arXiv.org/abs/2104.02349}{{\tt
  arXiv:2104.02349 [hep-ph]}}].

\bibitem{H1:2000kis}
{\bf H1} collaboration, C.~Adloff {\em et.~al.}, {\it {Elastic photoproduction
  of $J/\psi$ and $\Upsilon$ mesons at HERA}},
  \href{http://dx.doi.org/10.1016/S0370-2693(00)00530-X}{{\em Phys. Lett. B}
  {\bf 483} (2000) 23} [\href{http://arXiv.org/abs/hep-ex/0003020}{{\tt
  arXiv:hep-ex/0003020}}].

\bibitem{ZEUS:2002wfj}
{\bf ZEUS} collaboration, S.~Chekanov {\em et.~al.}, {\it {Exclusive
  photoproduction of $J/\psi$ mesons at HERA}},
  \href{http://dx.doi.org/10.1007/s10052-002-0953-7}{{\em Eur. Phys. J. C} {\bf
  24} (2002) 345} [\href{http://arXiv.org/abs/hep-ex/0201043}{{\tt
  arXiv:hep-ex/0201043}}].

\bibitem{LHCb:2014acg}
{\bf LHCb} collaboration, R.~Aaij {\em et.~al.}, {\it {Updated measurements of
  exclusive $J/\psi$ and $\psi$(2S) production cross-sections in pp collisions
  at $\sqrt{s}=7$ TeV}},
  \href{http://dx.doi.org/10.1088/0954-3899/41/5/055002}{{\em J. Phys. G} {\bf
  41} (2014) 055002} [\href{http://arXiv.org/abs/1401.3288}{{\tt
  arXiv:1401.3288 [hep-ex]}}].

\bibitem{LHCb:2018rcm}
{\bf LHCb} collaboration, R.~Aaij {\em et.~al.}, {\it {Central exclusive
  production of $J/\psi$ and $\psi(2S)$ mesons in $pp$ collisions at
  $\sqrt{s}=13~$TeV}},  \href{http://dx.doi.org/10.1007/JHEP10(2018)167}{{\em
  JHEP} {\bf 10} (2018) 167} [\href{http://arXiv.org/abs/1806.04079}{{\tt
  arXiv:1806.04079 [hep-ex]}}].

\bibitem{Ivanov:2004vd}
D.~Y. Ivanov, A.~Schafer, L.~Szymanowski and G.~Krasnikov, {\it {Exclusive
  photoproduction of a heavy vector meson in QCD}},
  \href{http://dx.doi.org/10.1140/epjc/s2004-01712-x}{{\em Eur. Phys. J. C}
  {\bf 34} (2004)~no.~3 297} [\href{http://arXiv.org/abs/hep-ph/0401131}{{\tt
  arXiv:hep-ph/0401131}}].
\newblock [Erratum: Eur.Phys.J.C 75, 75 (2015)].

\bibitem{Jones:2015nna}
S.~P. Jones, A.~D. Martin, M.~G. Ryskin and T.~Teubner, {\it {Exclusive
  $J/\psi$ and $\Upsilon$ photoproduction and the low $x$ gluon}},
  \href{http://dx.doi.org/10.1088/0954-3899/43/3/035002}{{\em J. Phys. G} {\bf
  43} (2016)~no.~3 035002} [\href{http://arXiv.org/abs/1507.06942}{{\tt
  arXiv:1507.06942 [hep-ph]}}].

\bibitem{Flett:2019nga}
C.~A. Flett, S.~P. Jones, A.~D. Martin, M.~G. Ryskin and T.~Teubner, {\it
  {Towards a determination of the low $x$ gluon via exclusive $J/\psi$
  production}},  \href{http://dx.doi.org/10.22323/1.352.0053}{{\em PoS} {\bf
  DIS2019} (2019) 053} [\href{http://arXiv.org/abs/1907.06471}{{\tt
  arXiv:1907.06471 [hep-ph]}}].

\bibitem{Flett:2019pux}
C.~A. Flett, S.~P. Jones, A.~D. Martin, M.~G. Ryskin and T.~Teubner, {\it {How
  to include exclusive $J/\psi$ production data in global PDF analyses}},
  \href{http://dx.doi.org/10.1103/PhysRevD.101.094011}{{\em Phys. Rev. D} {\bf
  101} (2020)~no.~9 094011} [\href{http://arXiv.org/abs/1908.08398}{{\tt
  arXiv:1908.08398 [hep-ph]}}].

\bibitem{Flett:2020duk}
C.~A. Flett, A.~D. Martin, M.~G. Ryskin and T.~Teubner, {\it {Very low $x$
  gluon density determined by LHCb exclusive $J/\psi$ data}},
  \href{http://dx.doi.org/10.1103/PhysRevD.102.114021}{{\em Phys. Rev. D} {\bf
  102} (2020) 114021} [\href{http://arXiv.org/abs/2006.13857}{{\tt
  arXiv:2006.13857 [hep-ph]}}].

\bibitem{Flett:2021xsl}
C.~A. Flett, {\em {Exclusive Observables to NLO and Low $x$ PDF Phenomenology
  at the LHC}}.
\newblock PhD thesis, U. Liverpool (main), 2021.

\bibitem{ALICE:2021gpt}
{\bf ALICE} collaboration, S.~Acharya {\em et.~al.}, {\it {Coherent $J/ \psi$
  and $\psi$' photoproduction at midrapidity in ultra-peripheral
  Pb\textendash{}Pb ~collisions at~$\sqrt {s_{NN}}$ = 5.02 TeV}},
  \href{http://dx.doi.org/10.1140/epjc/s10052-021-09437-6}{{\em Eur. Phys. J.
  C} {\bf 81} (2021)~no.~8 712} [\href{http://arXiv.org/abs/2101.04577}{{\tt
  arXiv:2101.04577 [nucl-ex]}}].

\bibitem{ALICE:2013wjo}
{\bf ALICE} collaboration, E.~Abbas {\em et.~al.}, {\it {Charmonium and
  $e^+e^-$ pair photoproduction at mid-rapidity in ultra-peripheral Pb-Pb
  collisions at $\sqrt{s_{\rm NN}}$=2.76 TeV}},
  \href{http://dx.doi.org/10.1140/epjc/s10052-013-2617-1}{{\em Eur. Phys. J. C}
  {\bf 73} (2013)~no.~11 2617} [\href{http://arXiv.org/abs/1305.1467}{{\tt
  arXiv:1305.1467 [nucl-ex]}}].

\bibitem{ALICE:2019tqa}
{\bf ALICE} collaboration, S.~Acharya {\em et.~al.}, {\it {Coherent J/$\psi$
  photoproduction at forward rapidity in ultra-peripheral Pb-Pb collisions at
  $\sqrt{s_{\rm{NN}}}=5.02$ TeV}},
  \href{http://dx.doi.org/10.1016/j.physletb.2019.134926}{{\em Phys. Lett. B}
  {\bf 798} (2019) 134926} [\href{http://arXiv.org/abs/1904.06272}{{\tt
  arXiv:1904.06272 [nucl-ex]}}].

\bibitem{ALICE:2012yye}
{\bf ALICE} collaboration, B.~Abelev {\em et.~al.}, {\it {Coherent $J/\psi$
  photoproduction in ultra-peripheral Pb-Pb collisions at $\sqrt{s_{NN}} =
  2.76$ TeV}},  \href{http://dx.doi.org/10.1016/j.physletb.2012.11.059}{{\em
  Phys. Lett. B} {\bf 718} (2013) 1273}
  [\href{http://arXiv.org/abs/1209.3715}{{\tt arXiv:1209.3715 [nucl-ex]}}].

\bibitem{CMS:2016itn}
{\bf CMS} collaboration, V.~Khachatryan {\em et.~al.}, {\it {Coherent $J/\psi$
  photoproduction in ultra-peripheral PbPb collisions at $\sqrt {s_{NN}} =$
  2.76 TeV with the CMS experiment}},
  \href{http://dx.doi.org/10.1016/j.physletb.2017.07.001}{{\em Phys. Lett. B}
  {\bf 772} (2017) 489} [\href{http://arXiv.org/abs/1605.06966}{{\tt
  arXiv:1605.06966 [nucl-ex]}}].

\bibitem{LHCb:2021bfl}
{\bf LHCb} collaboration, R.~Aaij {\em et.~al.}, {\it {Study of coherent
  $J/\psi$ production in lead-lead collisions at $\sqrt{s_{NN}} = 5 TeV$}},
  \href{http://arXiv.org/abs/2107.03223}{{\tt arXiv:2107.03223 [hep-ex]}}.

\bibitem{Brewer:2021kiv}
J.~Brewer, A.~Mazeliauskas and W.~van~der Schee in {\em {Opportunities of OO
  and pO collisions at the LHC}}, 3, 2021.
\newblock \href{http://arXiv.org/abs/2103.01939}{{\tt arXiv:2103.01939
  [hep-ph]}}.

\bibitem{Kovarik:2015cma}
K.~Kovarik {\em et.~al.}, {\it {nCTEQ15 - Global analysis of nuclear parton
  distributions with uncertainties in the CTEQ framework}},
  \href{http://dx.doi.org/10.1103/PhysRevD.93.085037}{{\em Phys. Rev. D} {\bf
  93} (2016)~no.~8 085037} [\href{http://arXiv.org/abs/1509.00792}{{\tt
  arXiv:1509.00792 [hep-ph]}}].

\bibitem{Eskola:2016oht}
K.~J. Eskola, P.~Paakkinen, H.~Paukkunen and C.~A. Salgado, {\it {EPPS16:
  Nuclear parton distributions with LHC data}},
  \href{http://dx.doi.org/10.1140/epjc/s10052-017-4725-9}{{\em Eur. Phys. J. C}
  {\bf 77} (2017)~no.~3 163} [\href{http://arXiv.org/abs/1612.05741}{{\tt
  arXiv:1612.05741 [hep-ph]}}].

\bibitem{AbdulKhalek:2020yuc}
R.~Abdul~Khalek, J.~J. Ethier, J.~Rojo and G.~van Weelden, {\it {nNNPDF2.0:
  quark flavor separation in nuclei from LHC data}},
  \href{http://dx.doi.org/10.1007/JHEP09(2020)183}{{\em JHEP} {\bf 09} (2020)
  183} [\href{http://arXiv.org/abs/2006.14629}{{\tt arXiv:2006.14629
  [hep-ph]}}].

\bibitem{Eskola:2021nhw}
K.~J. Eskola, P.~Paakkinen, H.~Paukkunen and C.~A. Salgado, {\it {EPPS21: A
  global QCD analysis of nuclear PDFs}},
  \href{http://arXiv.org/abs/2112.12462}{{\tt arXiv:2112.12462 [hep-ph]}}.

\bibitem{Khalek:2022zqe}
R.~A. Khalek, R.~Gauld, T.~Giani, E.~R. Nocera, T.~R. Rabemananjara and
  J.~Rojo, {\it {nNNPDF3.0: Evidence for a modified partonic structure in heavy
  nuclei}},  \href{http://arXiv.org/abs/2201.12363}{{\tt arXiv:2201.12363
  [hep-ph]}}.

\bibitem{Jones:2013pga}
S.~P. Jones, A.~D. Martin, M.~G. Ryskin and T.~Teubner, {\it {Probes of the
  small $x$ gluon via exclusive $J/\psi$ and $\Upsilon$ production at HERA and
  the LHC}},  \href{http://dx.doi.org/10.1007/JHEP11(2013)085}{{\em JHEP} {\bf
  11} (2013) 085} [\href{http://arXiv.org/abs/1307.7099}{{\tt arXiv:1307.7099
  [hep-ph]}}].

\bibitem{Chen:2019uit}
Z.-Q. Chen and C.-F. Qiao, {\it {NLO QCD corrections to exclusive
  electroproduction of quarkonium}},
  \href{http://dx.doi.org/10.1016/j.physletb.2019.134816}{{\em Phys. Lett. B}
  {\bf 797} (2019) 134816} [\href{http://arXiv.org/abs/1903.00171}{{\tt
  arXiv:1903.00171 [hep-ph]}}].
\newblock [Erratum: Phys.Lett. B, 135759 (2020)].

\bibitem{Flett:2021ghh}
C.~A. Flett, J.~A. Gracey, S.~P. Jones and T.~Teubner, {\it {Exclusive heavy
  vector meson electroproduction to NLO in collinear factorisation}},
  \href{http://dx.doi.org/10.1007/JHEP08(2021)150}{{\em JHEP} {\bf 08} (2021)
  150} [\href{http://arXiv.org/abs/2105.07657}{{\tt arXiv:2105.07657
  [hep-ph]}}].

\bibitem{Collins:1996fb}
J.~C. Collins, L.~Frankfurt and M.~Strikman, {\it {Factorization for hard
  exclusive electroproduction of mesons in QCD}},
  \href{http://dx.doi.org/10.1103/PhysRevD.56.2982}{{\em Phys. Rev. D} {\bf 56}
  (1997) 2982} [\href{http://arXiv.org/abs/hep-ph/9611433}{{\tt
  arXiv:hep-ph/9611433}}].

\bibitem{Diehl:2003ny}
M.~Diehl, {\it {Generalized parton distributions}},
  \href{http://dx.doi.org/10.1016/j.physrep.2003.08.002}{{\em Phys. Rept.} {\bf
  388} (2003) 41} [\href{http://arXiv.org/abs/hep-ph/0307382}{{\tt
  arXiv:hep-ph/0307382}}].

\bibitem{Shuvaev:1999ce}
A.~G. Shuvaev, K.~J. Golec-Biernat, A.~D. Martin and M.~G. Ryskin, {\it {Off
  diagonal distributions fixed by diagonal partons at small x and xi}},
  \href{http://dx.doi.org/10.1103/PhysRevD.60.014015}{{\em Phys. Rev. D} {\bf
  60} (1999) 014015} [\href{http://arXiv.org/abs/hep-ph/9902410}{{\tt
  arXiv:hep-ph/9902410}}].

\bibitem{Shuvaev:1999fm}
A.~Shuvaev, {\it {Solution of the off forward leading logarithmic evolution
  equation based on the Gegenbauer moments inversion}},
  \href{http://dx.doi.org/10.1103/PhysRevD.60.116005}{{\em Phys. Rev. D} {\bf
  60} (1999) 116005} [\href{http://arXiv.org/abs/hep-ph/9902318}{{\tt
  arXiv:hep-ph/9902318}}].

\bibitem{Freund:2002qf}
A.~Freund, M.~McDermott and M.~Strikman, {\it {Modeling generalized parton
  distributions to describe deeply virtual Compton scattering data}},
  \href{http://dx.doi.org/10.1103/PhysRevD.67.036001}{{\em Phys. Rev. D} {\bf
  67} (2003) 036001} [\href{http://arXiv.org/abs/hep-ph/0208160}{{\tt
  arXiv:hep-ph/0208160}}].

\bibitem{Martin:2008gqx}
A.~D. Martin, C.~Nockles, M.~G. Ryskin, A.~G. Shuvaev and T.~Teubner, {\it
  {Generalised parton distributions at small x}},
  \href{http://dx.doi.org/10.1140/epjc/s10052-009-1087-y}{{\em Eur. Phys. J. C}
  {\bf 63} (2009) 57} [\href{http://arXiv.org/abs/0812.3558}{{\tt
  arXiv:0812.3558 [hep-ph]}}].

\bibitem{Kumericki:2009uq}
K.~Kumeri\v{c}ki and D.~Mueller, {\it {Deeply virtual Compton scattering at
  small $x_B$ and the access to the GPD H}},
  \href{http://dx.doi.org/10.1016/j.nuclphysb.2010.07.015}{{\em Nucl. Phys. B}
  {\bf 841} (2010) 1} [\href{http://arXiv.org/abs/0904.0458}{{\tt
  arXiv:0904.0458 [hep-ph]}}].

\bibitem{Harland-Lang:2013xba}
L.~A. Harland-Lang, {\it {Simple form for the low-x generalized parton
  distributions in the skewed regime}},
  \href{http://dx.doi.org/10.1103/PhysRevD.88.034029}{{\em Phys. Rev. D} {\bf
  88} (2013)~no.~3 034029} [\href{http://arXiv.org/abs/1306.6661}{{\tt
  arXiv:1306.6661 [hep-ph]}}].

\bibitem{Constantinou:2020hdm}
M.~Constantinou {\em et.~al.}, {\it {Parton distributions and lattice-QCD
  calculations: Toward 3D structure}},
  \href{http://dx.doi.org/10.1016/j.ppnp.2021.103908}{{\em Prog. Part. Nucl.
  Phys.} {\bf 121} (2021) 103908} [\href{http://arXiv.org/abs/2006.08636}{{\tt
  arXiv:2006.08636 [hep-ph]}}].

\bibitem{Vidovic:1992ik}
M.~Vidovic, M.~Greiner, C.~Best and G.~Soff, {\it {Impact parameter dependence
  of the electromagnetic particle production in ultrarelativistic heavy ion
  collisions}},  \href{http://dx.doi.org/10.1103/PhysRevC.47.2308}{{\em Phys.
  Rev. C} {\bf 47} (1993) 2308}.

\bibitem{Guzey:2013taa}
V.~Guzey and M.~Zhalov, {\it {Rapidity and momentum transfer distributions of
  coherent $J/\psi$ photoproduction in ultraperipheral pPb collisions at the
  LHC}},  \href{http://dx.doi.org/10.1007/JHEP02(2014)046}{{\em JHEP} {\bf 02}
  (2014) 046} [\href{http://arXiv.org/abs/1307.6689}{{\tt arXiv:1307.6689
  [hep-ph]}}].

\bibitem{Dulat:2015mca}
S.~Dulat, T.-J. Hou, J.~Gao, M.~Guzzi, J.~Huston, P.~Nadolsky, J.~Pumplin,
  C.~Schmidt, D.~Stump and C.~P. Yuan, {\it {New parton distribution functions
  from a global analysis of quantum chromodynamics}},
  \href{http://dx.doi.org/10.1103/PhysRevD.93.033006}{{\em Phys. Rev. D} {\bf
  93} (2016)~no.~3 033006} [\href{http://arXiv.org/abs/1506.07443}{{\tt
  arXiv:1506.07443 [hep-ph]}}].

\bibitem{Drees:1988pp}
M.~Drees and D.~Zeppenfeld, {\it {Production of Supersymmetric Particles in
  Elastic $e p$ Collisions}},
  \href{http://dx.doi.org/10.1103/PhysRevD.39.2536}{{\em Phys. Rev. D} {\bf 39}
  (1989) 2536}.

\bibitem{Bertulani:2005ru}
C.~A. Bertulani, S.~R. Klein and J.~Nystrand, {\it {Physics of ultra-peripheral
  nuclear collisions}},
  \href{http://dx.doi.org/10.1146/annurev.nucl.55.090704.151526}{{\em Ann. Rev.
  Nucl. Part. Sci.} {\bf 55} (2005) 271}
  [\href{http://arXiv.org/abs/nucl-ex/0502005}{{\tt arXiv:nucl-ex/0502005}}].

\bibitem{ALICE:2021tyx}
{\bf ALICE} collaboration, S.~Acharya {\em et.~al.}, {\it {First measurement of
  the $|t|$-dependence of coherent $J/\psi$ photonuclear production}},
  \href{http://dx.doi.org/10.1016/j.physletb.2021.136280}{{\em Phys. Lett. B}
  {\bf 817} (2021) 136280} [\href{http://arXiv.org/abs/2101.04623}{{\tt
  arXiv:2101.04623 [nucl-ex]}}].

\bibitem{Klein:1999qj}
S.~Klein and J.~Nystrand, {\it {Exclusive vector meson production in
  relativistic heavy ion collisions}},
  \href{http://dx.doi.org/10.1103/PhysRevC.60.014903}{{\em Phys. Rev. C} {\bf
  60} (1999) 014903} [\href{http://arXiv.org/abs/hep-ph/9902259}{{\tt
  arXiv:hep-ph/9902259}}].

\bibitem{Woods:1954zz}
R.~D. Woods and D.~S. Saxon, {\it {Diffuse Surface Optical Model for
  Nucleon-Nuclei Scattering}},
  \href{http://dx.doi.org/10.1103/PhysRev.95.577}{{\em Phys. Rev.} {\bf 95}
  (1954) 577}.

\bibitem{DeVries:1987atn}
H.~De~Vries, C.~W. De~Jager and C.~De~Vries, {\it {Nuclear charge and
  magnetization density distribution parameters from elastic electron
  scattering}},  \href{http://dx.doi.org/10.1016/0092-640X(87)90013-1}{{\em
  Atom. Data Nucl. Data Tabl.} {\bf 36} (1987) 495}.

\bibitem{Helenius:2012wd}
I.~Helenius, K.~J. Eskola, H.~Honkanen and C.~A. Salgado, {\it
  {Impact-Parameter Dependent Nuclear Parton Distribution Functions: EPS09s and
  EKS98s and Their Applications in Nuclear Hard Processes}},
  \href{http://dx.doi.org/10.1007/JHEP07(2012)073}{{\em JHEP} {\bf 07} (2012)
  073} [\href{http://arXiv.org/abs/1205.5359}{{\tt arXiv:1205.5359 [hep-ph]}}].

\bibitem{H1:2013okq}
{\bf H1} collaboration, C.~Alexa {\em et.~al.}, {\it {Elastic and
  Proton-Dissociative Photoproduction of $J/\psi$ Mesons at HERA}},
  \href{http://dx.doi.org/10.1140/epjc/s10052-013-2466-y}{{\em Eur. Phys. J. C}
  {\bf 73} (2013)~no.~6 2466} [\href{http://arXiv.org/abs/1304.5162}{{\tt
  arXiv:1304.5162 [hep-ex]}}].

\bibitem{Khoze:2013dha}
V.~A. Khoze, A.~D. Martin and M.~G. Ryskin, {\it {Diffraction at the LHC}},
  \href{http://dx.doi.org/10.1140/epjc/s10052-013-2503-x}{{\em Eur. Phys. J. C}
  {\bf 73} (2013) 2503} [\href{http://arXiv.org/abs/1306.2149}{{\tt
  arXiv:1306.2149 [hep-ph]}}].

\bibitem{Jones:2015phd}
S.~Jones, {\em A study of exclusive processes to {NLO} and {Small-x} {PDFs}
  from LHC Data}.
\newblock PhD thesis, 2015.
\newblock Private communication.

\bibitem{Berger:1980ni}
E.~L. Berger and D.~L. Jones, {\it {Inelastic Photoproduction of $J/\psi$ and
  $\Upsilon$ by Gluons}},
  \href{http://dx.doi.org/10.1103/PhysRevD.23.1521}{{\em Phys. Rev. D} {\bf 23}
  (1981) 1521}.

\bibitem{Petrelli:1997ge}
A.~Petrelli, M.~Cacciari, M.~Greco, F.~Maltoni and M.~L. Mangano, {\it {NLO
  production and decay of quarkonium}},
  \href{http://dx.doi.org/10.1016/S0550-3213(97)00801-8}{{\em Nucl. Phys. B}
  {\bf 514} (1998) 245} [\href{http://arXiv.org/abs/hep-ph/9707223}{{\tt
  arXiv:hep-ph/9707223}}].

\bibitem{Bodwin:2002cfe}
G.~T. Bodwin and A.~Petrelli, {\it {Order-$v^4$ corrections to $S$-wave
  quarkonium decay}},  \href{http://dx.doi.org/10.1103/PhysRevD.66.094011}{{\em
  Phys. Rev. D} {\bf 66} (2002) 094011}
  [\href{http://arXiv.org/abs/hep-ph/0205210}{{\tt arXiv:hep-ph/0205210}}].
\newblock [Erratum: Phys.Rev.D 87, 039902 (2013)].

\bibitem{Braaten:2002fi}
E.~Braaten and J.~Lee, {\it {Exclusive Double Charmonium Production from $e^+
  e^-$ Annihilation into a Virtual Photon}},
  \href{http://dx.doi.org/10.1103/PhysRevD.72.099901}{{\em Phys. Rev. D} {\bf
  67} (2003) 054007} [\href{http://arXiv.org/abs/hep-ph/0211085}{{\tt
  arXiv:hep-ph/0211085}}].
\newblock [Erratum: Phys.Rev.D 72, 099901 (2005)].

\bibitem{Barbieri:1975ki}
R.~Barbieri, R.~Gatto, R.~Kogerler and Z.~Kunszt, {\it {Meson hyperfine
  splittings and leptonic decays}},
  \href{http://dx.doi.org/10.1016/0370-2693(75)90267-1}{{\em Phys. Lett. B}
  {\bf 57} (1975) 455}.

\bibitem{Bodwin:1994jh}
G.~T. Bodwin, E.~Braaten and G.~P. Lepage, {\it {Rigorous QCD analysis of
  inclusive annihilation and production of heavy quarkonium}},
  \href{http://dx.doi.org/10.1103/PhysRevD.55.5853}{{\em Phys. Rev. D} {\bf 51}
  (1995) 1125} [\href{http://arXiv.org/abs/hep-ph/9407339}{{\tt
  arXiv:hep-ph/9407339}}].
\newblock [Erratum: Phys.Rev.D 55, 5853 (1997)].

\bibitem{Beneke:1997jm}
M.~Beneke, A.~Signer and V.~A. Smirnov, {\it {Two loop correction to the
  leptonic decay of quarkonium}},
  \href{http://dx.doi.org/10.1103/PhysRevLett.80.2535}{{\em Phys. Rev. Lett.}
  {\bf 80} (1998) 2535} [\href{http://arXiv.org/abs/hep-ph/9712302}{{\tt
  arXiv:hep-ph/9712302}}].

\bibitem{Ji:1996nm}
X.-D. Ji, {\it {Deeply virtual Compton scattering}},
  \href{http://dx.doi.org/10.1103/PhysRevD.55.7114}{{\em Phys. Rev. D} {\bf 55}
  (1997) 7114} [\href{http://arXiv.org/abs/hep-ph/9609381}{{\tt
  arXiv:hep-ph/9609381}}].

\bibitem{Zyla:2020zbs}
{\bf Particle Data Group} collaboration, P.~Zyla {\em et.~al.}, {\it {Review of
  Particle Physics}},  \href{http://dx.doi.org/10.1093/ptep/ptaa104}{{\em PTEP}
  {\bf 2020} (2020)~no.~8 083C01}.

\bibitem{Buckley:2014ana}
A.~Buckley, J.~Ferrando, S.~Lloyd, K.~Nordstr\"om, B.~Page, M.~R\"ufenacht,
  M.~Sch\"onherr and G.~Watt, {\it {LHAPDF6: parton density access in the LHC
  precision era}},
  \href{http://dx.doi.org/10.1140/epjc/s10052-015-3318-8}{{\em Eur. Phys. J. C}
  {\bf 75} (2015) 132} [\href{http://arXiv.org/abs/1412.7420}{{\tt
  arXiv:1412.7420 [hep-ph]}}].

\bibitem{Florkowski:2010zz}
W.~Florkowski, {\em {Phenomenology of Ultra-Relativistic Heavy-Ion
  Collisions}}.
\newblock 3, 2010.

\bibitem{Zha:2018ywo}
W.~Zha, L.~Ruan, Z.~Tang, Z.~Xu and S.~Yang, {\it {Coherent lepton pair
  production in hadronic heavy ion collisions}},
  \href{http://dx.doi.org/10.1016/j.physletb.2018.04.006}{{\em Phys. Lett. B}
  {\bf 781} (2018) 182} [\href{http://arXiv.org/abs/1804.01813}{{\tt
  arXiv:1804.01813 [hep-ph]}}].

\bibitem{vonWeizsacker:1934nji}
C.~F. von Weizsacker, {\it {Radiation emitted in collisions of very fast
  electrons}},  \href{http://dx.doi.org/10.1007/BF01333110}{{\em Z. Phys.} {\bf
  88} (1934) 612}.

\bibitem{Jackson:1998nia}
J.~D. Jackson, {\em {Classical Electrodynamics}}.
\newblock Wiley, 1998.

\bibitem{Martin:1997wy}
A.~D. Martin and M.~G. Ryskin, {\it {The effect of off diagonal parton
  distributions in diffractive vector meson electroproduction}},
  \href{http://dx.doi.org/10.1103/PhysRevD.57.6692}{{\em Phys. Rev. D} {\bf 57}
  (1998) 6692} [\href{http://arXiv.org/abs/hep-ph/9711371}{{\tt
  arXiv:hep-ph/9711371}}].

\bibitem{Hoodbhoy:1996zg}
P.~Hoodbhoy, {\it {Wave function corrections and off forward gluon
  distributions in diffractive $J/\psi$ electroproduction}},
  \href{http://dx.doi.org/10.1103/PhysRevD.56.388}{{\em Phys. Rev. D} {\bf 56}
  (1997) 388} [\href{http://arXiv.org/abs/hep-ph/9611207}{{\tt
  arXiv:hep-ph/9611207}}].

\bibitem{Frankfurt:1997fj}
L.~Frankfurt, W.~Koepf and M.~Strikman, {\it {Diffractive heavy quarkonium
  photoproduction and electroproduction in QCD}},
  \href{http://dx.doi.org/10.1103/PhysRevD.57.512}{{\em Phys. Rev. D} {\bf 57}
  (1998) 512} [\href{http://arXiv.org/abs/hep-ph/9702216}{{\tt
  arXiv:hep-ph/9702216}}].

\bibitem{Lappi:2020ufv}
T.~Lappi, H.~M\"antysaari and J.~Penttala, {\it {Relativistic corrections to
  the vector meson light front wave function}},
  \href{http://dx.doi.org/10.1103/PhysRevD.102.054020}{{\em Phys. Rev. D} {\bf
  102} (2020)~no.~5 054020} [\href{http://arXiv.org/abs/2006.02830}{{\tt
  arXiv:2006.02830 [hep-ph]}}].

\end{thebibliography}\endgroup

\end{document}